\documentclass{aa}
\usepackage{soul} 
\usepackage{xcolor}
\usepackage[colorlinks, citecolor=blue, linkcolor=blue]{hyperref}

\usepackage{multirow}

\usepackage{graphicx}
\usepackage{subcaption}
\usepackage{txfonts}
\begin{document}

   \title{Unsupervised learning reveals a small number of galaxy formation pathways in IllustrisTNG}

   \author{C. Álvarez-Fernández\inst{1}\thanks{e-mail: calvarez@comillas.edu}
            \and H. Domínguez Sánchez \inst{2}
            \and J. Vega-Ferrero \inst{3,4}
            \and E. Angeloudi \inst{5,6}
            \and A. Ferré-Mateu \inst{5}
            \and D. Sotillo-Ramos \inst{7}
          }

   \institute{%
        \inst{1} Departamento de Métodos Cuantitativos, Santalucía Chair of Analytics for Education, Universidad Pontificia Comillas, C. de Alberto Aguilera, 23, 28015 Madrid, Spain \\
        \inst{2} Instituto de Física de Cantabria, Av. de los Castros, 39005 Santander, Cantabria, Spain\\
        \inst{3} Centro de Estudios de Física del Cosmos de Aragón (CEFCA), Plaza de San Juan, 1, E-44001 Teruel, Spain\\
        \inst{4} Deduce Data Solutions, C. Isabel Torres, 13, 39011 Santander, Cantabria, Spain\\
        \inst{5} Instituto de Astrofísica de Canarias, C. Vía Láctea s/n, E38205 - La Laguna, Tenerife, Spain\\
        \inst{6} Departamento de Astrofísica, Universidad de La Laguna, E-38205 La Laguna, Tenerife, Spain\\
        \inst{7} Departamento de Física Teórica, Universidad Autónoma de Madrid, Ciudad Universitaria de Cantoblanco, 28049 Madrid, Spain\\
    }

   %\date{Received September 15, 1996; accepted March 16, 1997}
    \date{Received XXX; accepted YYY}

 \abstract{Understanding the main drivers of galaxy formation and the connection between observable properties and evolutionary pathways is a central goal of extragalactic astronomy. Observations provide only single-epoch snapshots of galaxies at different stages of their development, whereas large-volume cosmological simulations allow us to trace the full formation histories of individual systems within a theoretical framework.}{Extracting physically meaningful information from these complex and high-dimensional datasets remains challenging. Our aim is to use machine-learning techniques to unveil physical properties and evolutionary pathways from the simulation datasets.}{In this work, we apply two unsupervised machine-learning techniques, dimensionality reduction and clustering, to identify galaxy evolution key parameters. We use uniform manifold approximation and projection (UMAP)  as our dimensionality reduction tool and $k$-means as our clustering method.} {We perform dimensionality reduction on the star formation histories (SFHs) of a sample of $\sim$10000 galaxies drawn from the TNG50 and TNG100 simulations. The resulting representation of the SFHs in a two-dimensional UMAP space reveals a clear ordering of galaxy mass assembly histories, naturally separating quenched and star-forming systems, as well as galaxies with significant ex~situ stellar mass contributions. Galaxy properties from simulation catalogs that are not used in the dimensionality reduction show strong correlations across the UMAP plane. Furthermore, using a $k$-means clustering analysis, we identify six main groups of galaxies with distinct evolutionary histories. These groups show systematic differences in stellar mass, morphology, angular momentum, star formation activity, and merger history, and occupy different regions of the SFR–mass and mass–size planes.}{Our results indicate that the diversity of galaxy formation histories in IllustrisTNG can be organized into a small number of characteristic evolutionary pathways linking galaxy accretion histories and present-day structural and star-forming properties.} 
% 5 {} token are mandatory


   \maketitle
%
%-------------------------------------------------------------------

\section{Introduction}

Understanding the physical processes that drive galaxy formation and determine the diversity of present-day galaxy properties remains a central goal of extragalactic astronomy. Galaxies display a wide range of structural, kinematic, and star-forming properties, which are thought to reflect different evolutionary pathways shaped by processes such as gas accretion, internal secular evolution, environmental effects, and galaxy mergers. Establishing how these mechanisms shape the observable properties of galaxies is therefore crucial to building a coherent picture of galaxy evolution. Observational studies have revealed several key scaling relations that encode important information about galaxy evolution, including the star formation rate (SFR) – stellar mass relation \citep[e.g.,][]{Brinchmann2004, Noeske2007, Whitaker2012} and the mass–size relation \citep[e.g.,][]{Buitrago2008, Buitrago2024, Shen2003, 2014ApJ...788...28V}. However, observations provide only instantaneous snapshots of galaxies at different stages of their life cycle, making it difficult to directly reconstruct their evolutionary histories.

Cosmological hydrodynamical simulations of galaxy formation offer a powerful complementary approach. Modern simulations follow galaxy formation within large cosmological volumes while modeling the baryonic processes that shape galaxy assembly. In particular, the IllustrisTNG project \citep{2018MNRAS.480.5113M, 2018MNRAS.477.1206N, 2018MNRAS.475..624N, 2018MNRAS.475..648P, 2018MNRAS.475..676S} provides a suite of simulations that simultaneously reproduce many observed galaxy properties while tracking the detailed formation histories of galaxies across cosmic time. These simulations therefore enable direct studies of galaxy evolutionary pathways. At the same time, the richness of the information they provide poses a significant challenge, as the resulting datasets are large, complex, and highly multidimensional.

Machine learning (ML) techniques have recently emerged as effective and robust tools for analyzing such datasets in astrophysics \citep[e.g.,][]{Baron2019, Villaescusa-Navarro2021, Villaescusa-Navarro2022}. In particular, ML methods can identify patterns and correlations in high-dimensional data that may not be easily detected using traditional analysis techniques. Supervised learning approaches have been widely used in observational data to classify galaxies and infer their physical properties, for example, in the context of morphological classification \citep[e.g.,][]{DS2018, DS2022, Vega-Ferrero2021}, size determination \citep{Vega-Ferrero2025}, or stellar population characterization \citep[e.g.,][]{Angeloudi2023,2024NatAs...8.1310A,Angeloudi2025,DS2026}. Although highly successful, these methods rely on predefined labels, which may bias the analysis toward existing observational classifications.

More recently, self-supervised and representation-learning approaches have been explored to extract information directly from imaging data without explicit labels \citep[e.g.,][]{Hayat2021, Martinez-Solaeche2024, 2021ApJ...921..177S, Stein2022, Vega-Ferrero2024}. These methods have shown great promise in capturing complex features in galaxy images and linking them to physical properties or evolutionary histories. However, many of these approaches focus on image-based representations, which do not explicitly encode the temporal dimension of galaxy evolution.

Unsupervised machine learning offers a complementary approach by allowing the intrinsic structure of the data to emerge without imposing predefined categories. Dimensionality reduction techniques, such as principal component analysis \citep[PCA;][]{pearson1901liii}, t-distributed stochastic neighbor encoding \citep[t-SNE;][]{vanDerMaaten2008} or uniform manifold approximation and projection \citep[UMAP;][]{McInnes2018}, are particularly well suited to exploring high-dimensional datasets by identifying the dominant modes of variation in the data and mapping them into a low-dimensional space. Clustering algorithms, such as $k$-means \citep{lloyd1982least, macqueen1967some, zbMATH03129892} can then be used to identify groups of objects with similar properties within this reduced representation.

In this work, we build on these ideas by applying a combination of dimensionality reduction and clustering techniques to the star formation histories (SFHs) of galaxies in the IllustrisTNG simulations. In contrast to previous studies that focus primarily on imaging data, our approach leverages the full temporal information encoded in the SFHs to identify evolutionary pathways in a fully data-driven manner. We analyze the SFHs of a sample of $\sim$10000 galaxies drawn from the TNG50 and TNG100 simulations. Using the UMAP algorithm to project the SFHs into a two-dimensional space, followed by a $k$-means clustering analysis, we identify groups of galaxies with similar evolutionary histories.

We then investigate how these groups relate to present-day observable galaxy properties, including morphology, stellar mass, size, star formation activity, angular momentum, and merger history. In particular, we explore how the different populations populate key scaling relations such as the star formation rate–stellar mass and mass–size relations. This approach provides a data-driven framework for linking galaxy assembly histories with the observable diversity of galaxies and identifying the dominant formation pathways present in cosmological simulations.

The paper is structured as follows. Section \ref{TNG100} presents the IllustrisTNG simulations and the catalogs used. Section \ref{Method} describes the methodology. Sections \ref{UMAP}  and \ref{Cluster} show the results of dimensionality reduction and clustering, respectively. Finally, Section \ref{Conclusions} summarizes our findings.

\section{Data: IllustrisTNG simulations}
\label{TNG100}

    \begin{figure*}
    \includegraphics[width=0.33\linewidth]{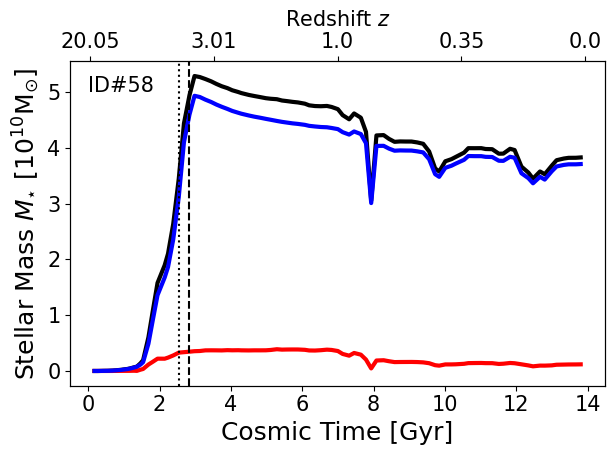}
    \includegraphics[width=0.33\linewidth]{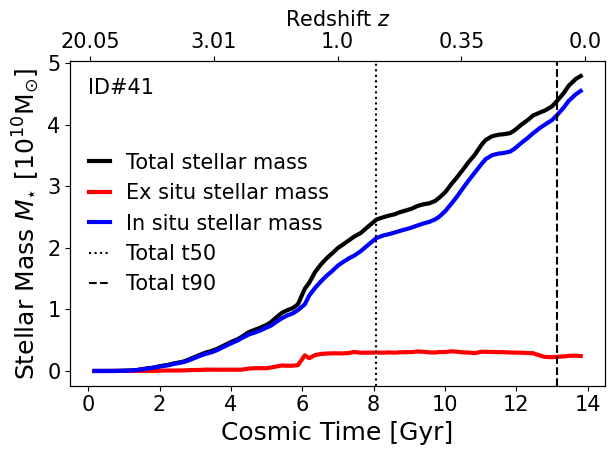}
    \includegraphics[width=0.33\linewidth]{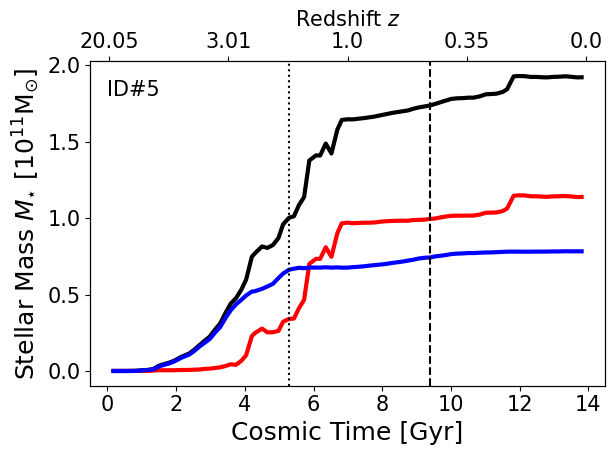}
    \caption{Star formation histories of three galaxies \#58 (left), \#41 (center), and \#5 (right), from the Stellar Assembly catalog (Illustris TNG100). We display stellar mass $M_{\star}$ [$\mathrm{M}_{\odot}$] over time [Gyr], presenting the in~situ stellar mass (blue), the ex~situ stellar mass (red) and the total stellar mass (black). Vertical lines are plotted at times where galaxies reach 50\% (dotted) and 90\% (dashed) of their maximum total stellar mass ($t_{50}$ and $t_{90}$).}
        \label{fig:SFH}
     \end{figure*} 

    In this work, we use data from the IllustrisTNG project\footnote{\url{https://www.tng-project.org/}}, in particular, from the TNG100 and TNG50 volumes.  IllustrisTNG is a suite of large-volume, cosmological, gravo-magnetohydrodynamical simulations run with the moving-mesh code \textsc{Arepo}\footnote{\url{https://arepo-code.org/}}. The details of \textsc{Arepo} code can be found in \cite{springel2010pur}, and the physical galaxy formation models are described in \cite{2018MNRAS.473.4077P,2017MNRAS.465.3291W}. IllustrisTNG includes a comprehensive model for galaxy formation physics, and each TNG simulation self-consistently solves for the coupled evolution of dark matter, cosmic gas, luminous stars, and supermassive black holes from the early times to the present day, $z=0$. The IllustrisTNG project is the successor of the Illustris simulation. It uses an updated ``next generation'' galaxy formation model which includes both new physics as well as refinements to the original Illustris model. The project encompasses three primary volumes (TNG50, TNG100, and TNG300) comprising 18 distinct simulations. They have different volume sizes and particle resolution. TNG50 has a volume side of 50 comoving Mpc and the highest particle resolution ($m_{baryon} = 8.5\times10^4 \, \mathrm{M}_{\odot}$), which is necessary for detailed galaxy modeling. TNG100 has a larger volume but lower particule resolution (100 Mpc volume, $m_{baryon} = 1.4\times10^6 \, \mathrm{M}_{\odot}$). Finally, TNG300 has the largest volume but the lowest particle resolution, which is useful for studying large-scale processes, but will not be used in this work.

    Each of the runs is accompanied by halo/subhalo catalogs, merger trees, lower-resolution and dark-matter-only counterparts, all available with 100 snapshots. More details on the simulations can be found in the release paper \cite{2019ComAC...6....2N} and references therein (e.g., \citealt{2018MNRAS.480.5113M, 2018MNRAS.477.1206N, 2018MNRAS.475..624N, 2019MNRAS.490.3234N, Pillepich1,  2018MNRAS.475..648P, 2018MNRAS.473.4077P, Pillepich_2019, 2018MNRAS.475..676S, 2017MNRAS.465.3291W}).

    In this work, we combine the TNG50 and TNG100 simulations to increase the sample size and the dynamic range of the galaxies. Although the two simulations have different mass resolution, several works have combined them in their analysis \citep{2026arXiv260912339C,Eisert2024,Engler2021,Joshi2020,2018MNRAS.475..648P}. To keep track of possible biases or significant differences, the two samples are separated or represented using different symbols in figures and tables. Finally, we performed an additional test using the TNG100 sample alone, showing minor differences that do not affect the main results presented in this work (see Figures~\ref{fig:Umapmixed} and \ref{fig:SFHCluster100}).

    In this work, we take advantage of the following public catalogs from IllustrisTNG: Stellar Assembly, Stellar Circularities and Angular Momenta, Galaxy Morphologies, Merger History, Star Formation Rate, and Stellar Projected Sizes, which we describe below.

    \subsection{Stellar assembly catalog}
    This is the starting point for our analysis and contains information about how galaxies grow in mass across  100 snapshots, focusing on in~situ vs. ex~situ stellar mass growth, i.e., formed within the main progenitor branch of a galaxy or accreted. The main progenitor of each subhalo is defined as the one with the most massive history behind it. Details can be found in \citep{Rodriguez_Gomez_2015,Rodriguez_Gomez_2016,2017MNRAS.467.3083R}. We use the following fields of the catalog, where all masses are measured in $\mathrm{M}_{\odot}$: 
        \begin{itemize}
            \item Snapshot: Indexed from 0 to 99, representing discrete time steps from the early universe ($ \approx 20$) to the present day ($z=0$).
            \item Galaxy ID: Integer number used to identify each galaxy. It is called SubfindID or SubhaloID in other catalogs.
            \item Stellar mass in~situ: The stellar mass ($M_{\star}$) formed within the galaxy.
            \item Stellar mass ex~situ: The stellar mass ($M_{\star}$) formed outside the galaxy.
            \item Total stellar mass: The sum of the in~situ and ex~situ masses.
        \end{itemize}

    Our TNG100 initial galaxy sample consists of 6503 different galaxies\footnote{The precise term used in the catalogs is subhalo, but we will use both terms indistinctly.} at $z=0$ with a stellar mass above $10^{10} \, \mathrm{M}_{\odot}$. Every galaxy is tracked in at most 100 snapshots, from the early Universe ($z \approx 20$, snapshot 0) to the present day ($z=0$, snapshot 99). We filtered out galaxies with less than ten nonzero values (less than 1\%) to eliminate the influence of galaxies with extremely short star formation histories. The final sample for TNG100 comprises 6481 galaxies from TNG100.

    For TNG50, the initial number of galaxies is 3238 and 3167 after filtering (the same process as for TNG100 explained above) with a minimum stellar mass of $10^9 \, \mathrm{M}_{\odot}$. In this work, we merge the data from TNG50 and TNG100 into a single dataset consisting of 9648 galaxies with 100 snapshots for each one.
    
    Regarding missing data, in the total dataset (TNG50 and TNG100 together) there are 1130 galaxies with missing snapshots ($\sim$ 10\% of the galaxies), but the total number of missing data is only 1260 ($\sim$ 0.1\% of the data): the majority of incomplete galaxies have only one missing value, the maximum number of gaps is four, and most importantly, there are no consecutive gaps. Therefore, we expect that linear interpolation for these data has a negligible effect on the overall SFH shapes. In addition, for some subhalos we inserted zero values at the beginning when needed (because not all galaxies are formed at snapshot zero) to complete the 100 snapshots per galaxy.

    The general purpose of this work is to group galaxies with similar SFHs to study their characteristics and to shed light on galaxy formation pathways.  Figure \ref{fig:SFH} displays an example of three distinct SFHs of galaxies with different stellar masses at $z=0$. For each we calculate the formation times $t_{50}$ and $t_{90}$, which measure the amount of time that a galaxy needs to reach 50\% or 90\% of its maximum stellar mass (starting from the time when the galaxy forms), respectively. There are clear differences between the three SFHs, with the left one showing a rapid increase in mass and an early quenching of the star formation, with very small ex~situ contribution; the middle one showing a constant increase of in~situ stellar mass with small ex~situ contribution; and the right one showing an episode of star formation, both with a significant ex~situ contribution (major merger). 

   \subsection{Mergers catalog}
   
    This catalog contains information about the merger history of all subhalos. Mergers are divided into three categories: major mergers (stellar mass ratio $>$ 1/4), minor mergers (stellar mass ratio between 1/10 and 1/4), and all mergers (any stellar mass ratio). Complete definitions and details can be found in \cite{2023MNRAS.519.2199E} and \cite{2017MNRAS.467.3083R}. We use the following fields:

\begin{itemize}
            \item MeanGasFraction: The mean ``cold'' (i.e., star-forming) gas fraction of all the objects that have merged with the galaxy in question, weighted by their maximum stellar masses.
            \item MeanMassRatio: The mean stellar mass ratio of all the mergers that a galaxy has undergone, weighted by the maximum stellar mass of the secondary progenitors.
            \item MeanRedshift: The mean redshift of all the mergers that a galaxy has undergone, weighted by the maximum stellar mass of the secondary progenitors.
        \end{itemize}

\subsection{Stellar circularities, angular momenta, and axis ratios catalogs}

This catalog includes information about these quantities, calculated after alignment with the total angular momentum vector of the stars within 10 times the stellar half-mass radius with respect to the particle
with the lowest gravitational potential of the subhalo. The second and third fields of the catalog are based on the distribution of the circularity parameter $\epsilon$, ranging from -1 to 1, being 0 a radial orbit and 1 a circular co-rotating orbit. First, the system is rotated so that the z-axis is aligned with the angular momentum vector; then, for every stellar particle with specific angular momentum, $J_z$, the magnitude $\frac{J_z}{J(E)}$ is calculated, being $J(E)$ the maximum angular momentum of the stellar particles at positions between 50 before and 50 after the particle in question in a list where the stellar particles are sorted by their binding energy. We analyze the values in snapshot 99 ($z=0$). The full description of the variables can be found in \cite{2015ApJ...804L..40G}, of which we use:
        \begin{itemize}
            \item SpecificAngMom: Specific Angular Momentum of the stars ($j_{\star}$), in km s$^{-1} \times$ kpc.
            \item CircAbove07Frac: The fractional mass of stars with $\epsilon > 0.7$, commonly defined as ``disk'' stars, i.e., those with significant (positive) rotational support.
            \item CircTwiceBelow0Frac: The fractional mass of stars with $\epsilon < 0$, multiplied by two, commonly defined as the ``bulge'' stars.
        \end{itemize}

\subsection{Morphological catalog}

This catalog contains the probabilities of galaxies having a certain visual-like morphology, as determined by a deep-learning algorithm. The algorithm was trained with labeled observational data\footnote{Despite the limitations of the supervised methods, they are still vital to connect the groups found by unsupervised techniques with observational data.} from the Sloan Digital Sky Survey (SDSS, references can be found in  \citealt{2006AJ....131.2332G, 2000AJ....120.1579Y}) and applied to mock SDSS-like images from the IllustrisTNG simulation. For complete definitions of each value, see \cite{Huertas_Company_2019} and \cite{Varma_2021}. We  use the classifications at $z=0$, and for this particular analysis, we only use the TNG100 catalog, as the TNG50 morphological classifications are not available for $z=0$ but for $z=0.5$. We use the following fields for all TNG100 galaxies at $z=0$.

        \begin{itemize}
            \item $P_{\text{Late}}$: Probability that the galaxy is a  late-type galaxy (LTG) rather than early-type (ETG).
            \item $P_{\text{S0}}$: Probability that the galaxy is S0 rather than elliptical (valid only for galaxies in which $ P_{\text{Late}} < 0.5$, i.e., likely ETG).
           % \item Sigma late, uncertainty in $P_{\text{Late}}$.
           % \item Sigma S0, uncertainty in $P_{S0}$.
        \end{itemize}

\subsection{Star formation rate catalog}

This is a catalog of time-averaged star formation rates (SFR) of galaxies. The SFRs are derived from stellar particles produced over different time spans, using their initial mass at birth. For complete definitions and descriptions of the calculation process for each value, see \cite{Donnari_2019} and \cite{Pillepich_2019}. In this project, we use the following:

\begin{itemize}
    \item SFR-MsunPerYrs-in-InRad-50Myrs: Named SFR50 here, it is the galaxy's SFR including stars within a 3D aperture of two times the stellar half-mass radius, averaged across the last 50 Myr.
\end{itemize}

\subsection{Stellar projected sizes catalog}
The  ``Stellar Projected Sizes'' is a catalog of stellar sizes  in 2D projection. The term ``size'' or ``effective radius'' for a galaxy can be defined in different ways. The details of the calculations can be found in \cite{2018MNRAS.474.3976G}, and in this project, we use this particular definition:

\begin{itemize}
    \item Half-mass radius: Radius that contains half of the stellar mass of the galaxy. The catalog gives some projection options, we chose the projection along the box z-axis, suggested there for a ``random'' projection as there is no preferred direction to the simulation.
\end{itemize}

\section{Methodology: dimensionality reduction and clustering}
\label{Method}

\begin{figure}
        \centering
        \includegraphics[width=1.00 \linewidth]{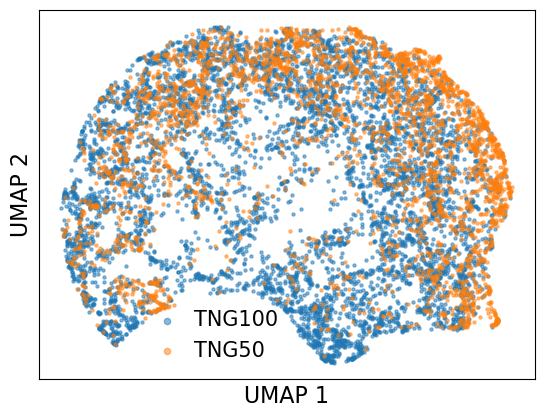}
        \includegraphics[width=1.00\linewidth]{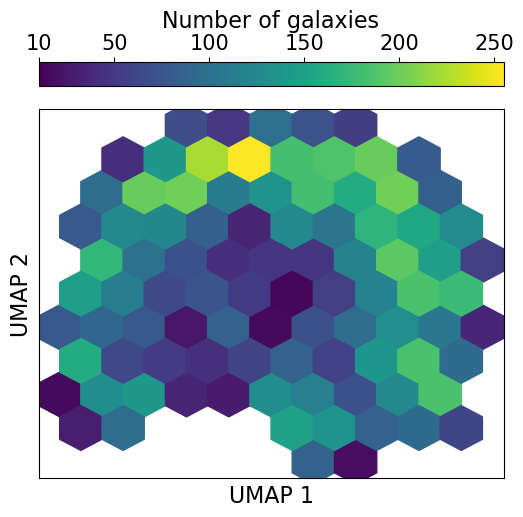}
\caption{Galaxy density plots, both individual and binned. Top: location of individual galaxies in the UMAP parametric space using two arbitrary dimensions UMAP 1 and UMAP 2. TNG100 and TNG50 are shown as blue and orange points, respectively. Bottom: Same UMAP plane showing galaxies binned in a 10 $\times$ 10 grid, the color code represents the number of galaxies in each bin combining TNG100 and TNG50 galaxies (we only represent bins with at least ten galaxies).}
\label{fig:UMAP}
\end{figure}

   To shed light on the galaxy mass assembly of TNG galaxies, we use two types of unsupervised learning techniques: dimensionality reduction and clustering. The use of dimensionality reduction is a key step in our analysis to extract information from high dimensional data (100 snapshots for roughly 10000 galaxies) that can be displayed in a 2D or 3D plot, simplifying further analysis without losing relevant information. We have explored three different techniques for dimensionality reduction: UMAP \citep{McInnes2018}, a nonlinear dimensionality reduction technique based on manifold learning and topological data analysis; PCA \citep{pearson1901liii}, a classical linear algebra based technique; and t-SNE \citep{vanDerMaaten2008}, a nonlinear stochastic method.
    
    After several tests, UMAP led to more robust results when combined with the subsequent cluster analysis, so we use it as our primary visualization tool. For simplicity, the analysis using PCA and t-SNE is not included in the main body of the manuscript; however, we found that the main trends are not sensitive to the choice of dimensionality reduction algorithm. We show some examples in the Appendix Fig \ref{fig:2DSFH}.

    Once the dimensions of the input data  have been reduced without losing substantial information, we use clustering techniques to find groups of similar elements in sets of unlabeled data. This will let us identify galaxies that share similar SFHs to then study if their observed properties are also similar. Clustering can be approached using two main methodologies: hierarchical and non-hierarchical methods (although there also exist hybrid and alternative methods). We have explored two clustering methods: $k$-means \citep{lloyd1982least, macqueen1967some, zbMATH03129892}, and hierarchical agglomerative clustering with euclidean distance and Ward linkage criterion \citep{Ward}. Throughout this work, we adopt the $k$-means results as a reference because it is computationally efficient and the results are robust, although the results given by the other methods are fully compatible, as shown in the Appendix Figure \ref{fig:ClusterWard}.

\section{Dimensionality reduction: UMAP}
\label{UMAP}

In the following sections, we use UMAP. It has been chosen because of its generality in capturing very general information in high-dimensionality data and its relatively low computational cost. 

We apply UMAP to the stellar mass assembly catalog described in Section \ref{TNG100} for the dataset combining the TNG100 and TNG50 catalogs. Each galaxy has an in~situ and ex~situ mass assembly, both with a length of 100 snapshots, so we concatenate both histories (first in~situ, then ex~situ) in a vector of length 200 for each galaxy. Consequently, the reduction process uses an input matrix of 9648 x 200 and gives as output a matrix of 9648 x 2 that can be plotted in a 2D scatter plot.  Before the dimensionality reduction, we  normalize the SFHs to the maximum total stellar mass of each galaxy to avoid the algorithm focusing on the stellar mass\footnote{Normalizing by the maximum mass provides a better physical representation of evolutionary pathways (e.g., preserving the signature of mass stripping) than normalizing by final mass.}. This allows finding relations among galaxies with different masses and guarantees that if galaxies of the same cluster have a similar mass, it is only driven by their evolutionary paths.

We use UMAP with its default parameters. The number of neighbors, set to 15, controls the balance between local and global structure. The minimum distance parameter, set to 0.1 by default\footnote{These values are the default values suggested in the software package documentation for general-purpose use.}, determines how closely points can be packed in the low-dimensional space. We set the number of components to 2 to obtain a two-dimensional representation. Although UMAP supports a wide range of distance metrics, for simplicity we use the default Euclidean metric, which provides results that are sufficiently interpretable for our purposes.

Figure \ref{fig:UMAP} shows the result of applying UMAP dimensionality reduction to the mass normalized SFHs vectors of each galaxy. Each point represents a galaxy, summarized in a 2D arbitrary vector (UMAP1, UMAP2), color-coded according to their simulation origin (TNG50 or TNG100).  Galaxies are not uniformly distributed, but there are regions where galaxies are more common. There is a significant overlap between TNG50 and TNG100 galaxies, supporting our strategy of combining data from both simulations together. However, given their differences in mass resolution and volume, the lower part of the UMAP is underpopulated by TNG50 galaxies. We will see in the next sections that this region precisely corresponds to the more massive galaxies. %(see Figure \ref{umapgen}). 

Figure \ref{fig:UMAP} also shows a binned plot color-coded by the number of galaxies included in each hexagonal region of the UMAP plane,  highlighting that some regions of the UMAP are more populated than others. In the next sections, we study how these regions are related to fundamental properties of galaxies such as mass, formation timescales, and merger rate (Section \ref{UMAP_SFH}) or other quantities unrelated to the information contained in the SFHs vectors (Section \ref{UMAP_cats}).

\subsection{Star formation histories in the UMAP plane}

\label{UMAP_SFH}

\begin{figure*}
\centering
\includegraphics[width=1.00 \textwidth]{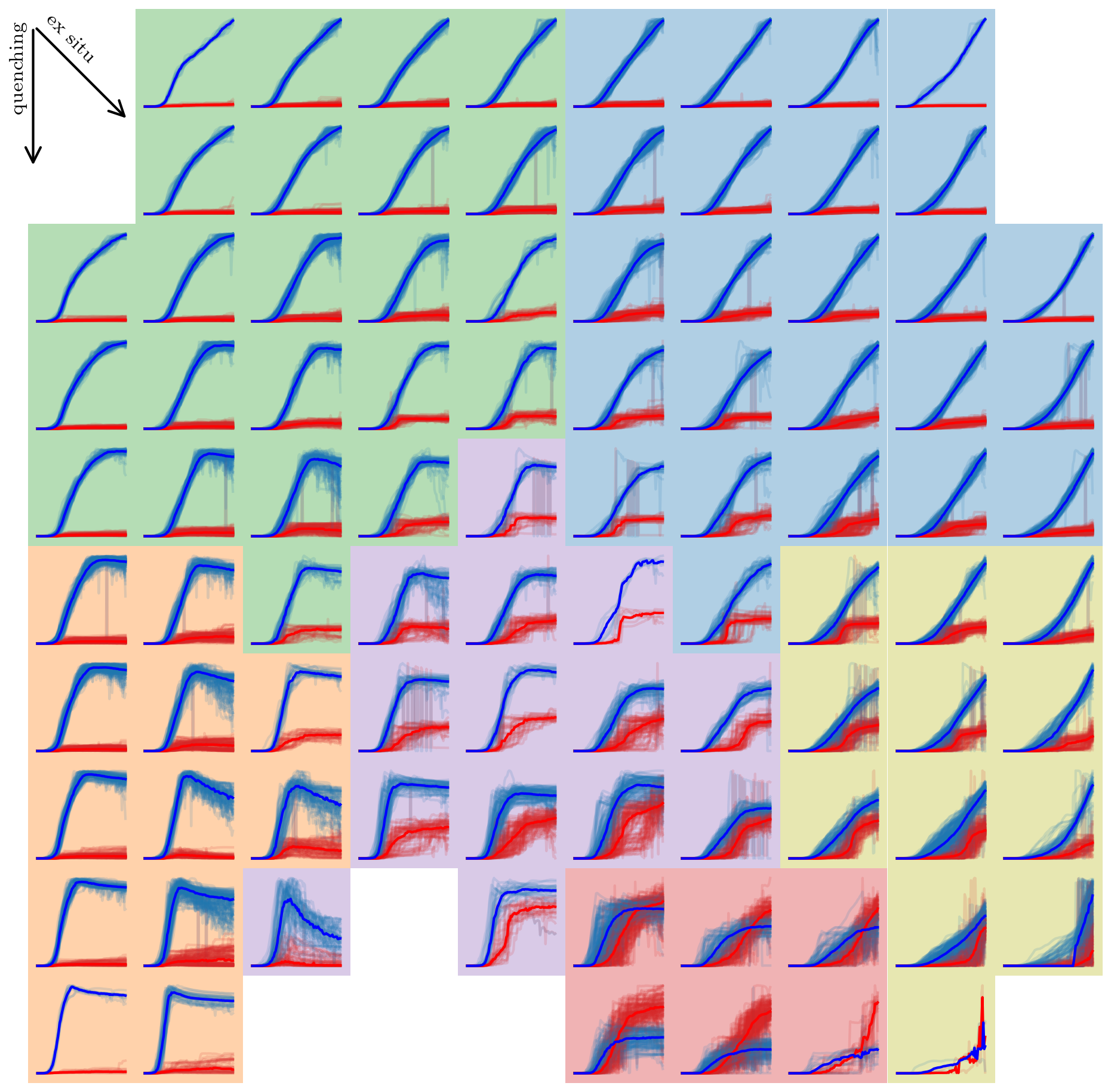}
\caption{\label{UMAPSFH} Star formation histories in UMAP representation. UMAP plot is divided using a 10 x 10 grid (as in Figure \ref{fig:UMAP}). For each cell in the grid, the SFHs (stellar mass vs lookbacktime, see figure \ref{fig:SFH}) of all the galaxies that lie in the cell are overlapped (ex~situ in red color and in~situ in blue color). Star formation histories are normalized to the maximum total stellar mass and median SFHs are shown as a thick line. Arrows are aligned with the main gradients in the plot, the vertical one corresponds to the degree of star formation quenching and the diagonal one measures the weight of the ex~situ component in the total stellar mass. The color code corresponds to the most frequent galaxy $k$-means cluster from Section \ref{Cluster}.}
\end{figure*}

The data used for dimensionality reduction are the star formation histories (SFHs) of galaxies. As an initial sanity check, we project this information onto the UMAP plane. Figure~\ref{UMAPSFH} displays, for each bin, the SFHs of the galaxies within that bin, along with the median SFH. A clear correlation emerges between the position in the UMAP space and the shape of the SFHs. Galaxies located in the upper region of the plot tend to exhibit steady SFHs with minimal ex~situ mass contributions. Those in the bottom-left corner are dominated by early and fast in~situ star formation, although some may have some ex~situ component. These galaxies typically show signs of quenching, with star formation being suppressed - and even showing apparent mass loss, compatible with interactions or tidal stripping by other galaxies. Lastly, galaxies in the bottom-right region are characterized by substantial ex~situ mass contributions and correspond to systems that have experienced major mergers in their lifetime. Although the UMAP axes do not have an intrinsic physical meaning, we show in Figure~\ref{UMAPSFH} vectors aligned with the direction of quenching and  ex~situ gradients. For visualization purposes, we added a color code based on the clustering presented in the following Section \ref{Cluster} (the background of each bin is plotted using the color of the most frequent cluster in it). 

\begin{figure*}
    \begin{subfigure}{0.33\textwidth}
    \centering
    \includegraphics[width=1.0\linewidth]{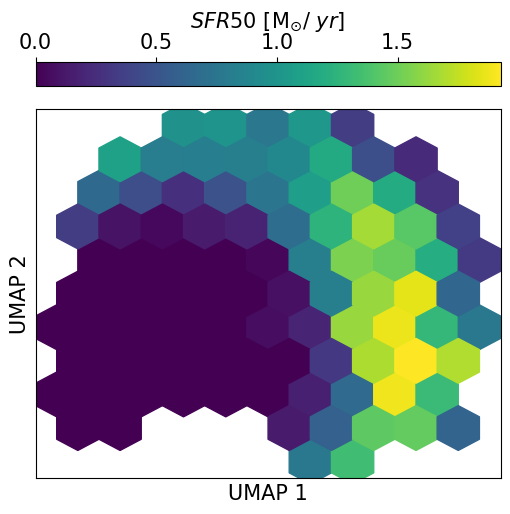}
         \caption{SFR across the last 50 Myr at $z=0$.}
         \label{fig:umapgen_a}
    \end{subfigure}
    \begin{subfigure}{0.33\textwidth}
    \centering
     \includegraphics[width=1.0\linewidth]{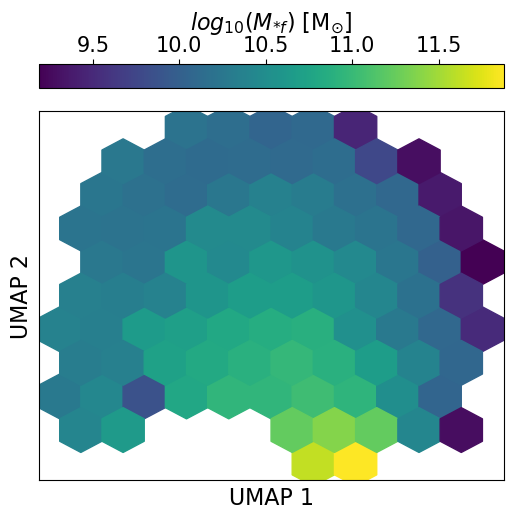}
         \caption{Final stellar mass $\log_{10} M_{\star f} [\mathrm{M}_{\odot}]$.}
         \label{fig:umapgen_b}    
    \end{subfigure}
    \begin{subfigure}{0.33\textwidth}
    \centering
     \includegraphics[width=1.0\linewidth]{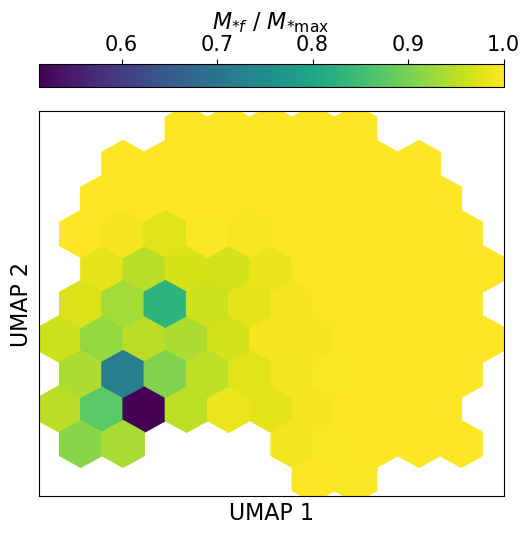}
         \caption{Final stellar mass $M_{\star f}$ / max mass $M_{\star \max}$.}
         \label{fig:umapgen_c}      
    \end{subfigure}
    \begin{subfigure}{0.33\textwidth}
    \centering
     \includegraphics[width=1.0\linewidth]{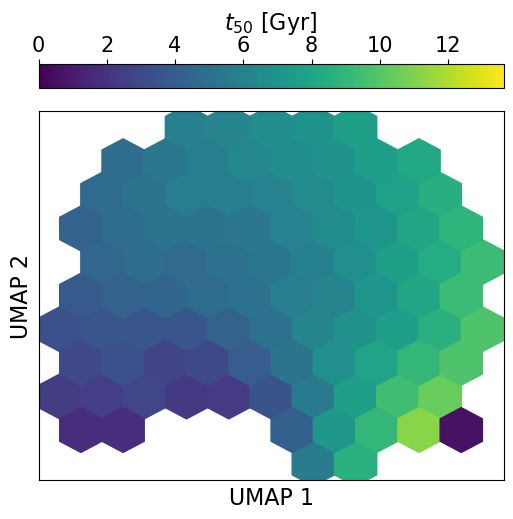}
         \caption{Formation time $t_{50}$ [Gyr].}
         \label{fig:umapgen_d}    
    \end{subfigure}
    \begin{subfigure}{0.33\textwidth}
    \centering
      \includegraphics[width=1.0\linewidth]{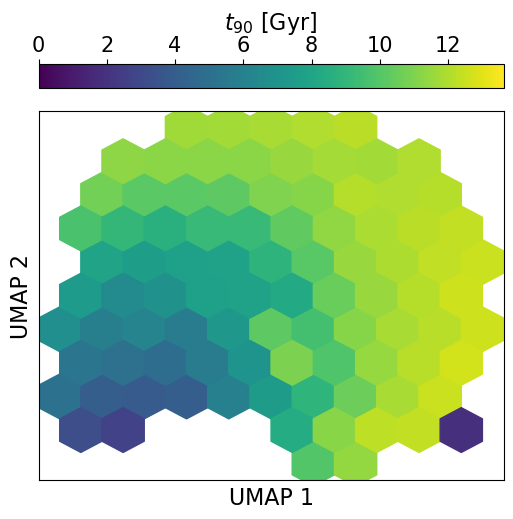}
         \caption{Formation time $t_{90}$ [Gyr].}
         \label{fig:umapgen_e}   
    \end{subfigure}
    \begin{subfigure}{0.33\textwidth}
    \centering
     \includegraphics[width=1.0\linewidth]{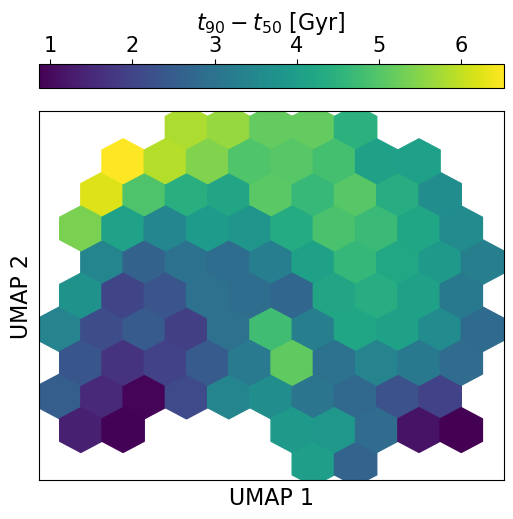}
         \caption{Formation time $t_{90}-t_{50}$ [Gyr].}
         \label{fig:umapgen_f}   
    \end{subfigure}
        \caption{Physical parameters (Star formation rate, mass, and formation times $t_{50}$ and $t_{90}$) displayed using the UMAP representation and arbitrary dimensions UMAP 1 and UMAP 2. We only represent bins with al least 10 galaxies.}
        \label{umapgen}
\end{figure*}

 Figure \ref{umapgen} displays the distribution of basic physical parameters directly related to SFHs (i.e., mass and formation times) in the UMAP plane. In Figure~\ref{fig:umapgen_a}, we also show the star formation within a 3D aperture of two times the stellar half-mass radius, averaged across the last 50 Myr. There is a clear gradient of SFR within the UMAP plane, with the galaxies showing the lowest SFR located in the bottom left, consistent with what is observed in Figure~\ref{UMAPSFH}. Galaxies located near the bottom right corner of the plot (those where the ex~situ component is significant) exhibit  diverse behaviors, suggesting that mergers can both enhance or quench star-formation processes. Figure~\ref{fig:umapgen_b} displays the total final stellar mass. It is important to remind the reader that SFHs are normalized to $M_{\star\max}$, so the UMAP representation does not use mass information as a parameter, but only the SFHs shape. Despite this normalization, we observe a smooth mass gradient from top to bottom. This means that there is a clear dependence between the total stellar mass and the formation pathways of galaxies. Figure~\ref{fig:umapgen_c} displays the ratio between the final total mass ($M_{\star f}$) and the maximum total mass ($M_{\star\max}$). Galaxies that have lost a significant amount of mass (i.e., $M_{\star f} / M_{\star \max} < 1$) lie near the bottom left corner, as suggested by  Figure \ref{UMAPSFH}. 
 
The bottom panels, Figures \ref{fig:umapgen_d}, \ref{fig:umapgen_e}, and \ref{fig:umapgen_f} display the formation timescales $t_{50}$, $t_{90}$, and their difference $t_{90} - t_{50}$. Following some works like \cite{Ferre-Mateu2023} and \cite{10.1093/mnras/stad2973}, we use these timescales to parametrize in a simple way the build up of stellar mass of galaxies. In particular, $t_{50}$ can be used as a proxy for the age of the galaxy and $t_{90}$ indicates the time of quenching.

A clear gradient in $t_{50}$ is observed, increasing from the bottom left toward the right side of the UMAP plane. The $t_{90}$ values are high both on the right side and in the upper region of the map. As a result, the upper left area shows large $t_{90} - t_{50}$ values, indicating galaxies whose mass assembly is slow. In contrast, the lower left region is populated by galaxies that formed rapidly, consistent with the SFHs presented in Figure~\ref{UMAPSFH}.

\begin{figure*}
\begin{subfigure}{.33\textwidth}
  \centering
  \includegraphics[width=1.0\linewidth]{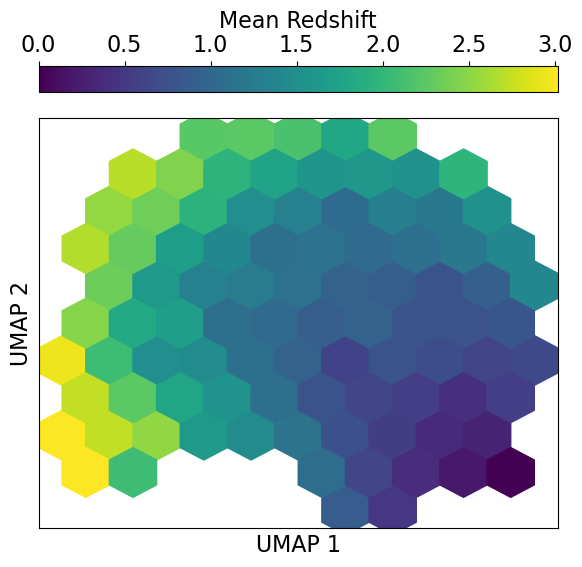}
  \caption{Mean Redshift}
  \label{fig:MergerUMAP7}
\end{subfigure}%
\begin{subfigure}{.33\textwidth}
  \centering
  \includegraphics[width=1.0\linewidth]{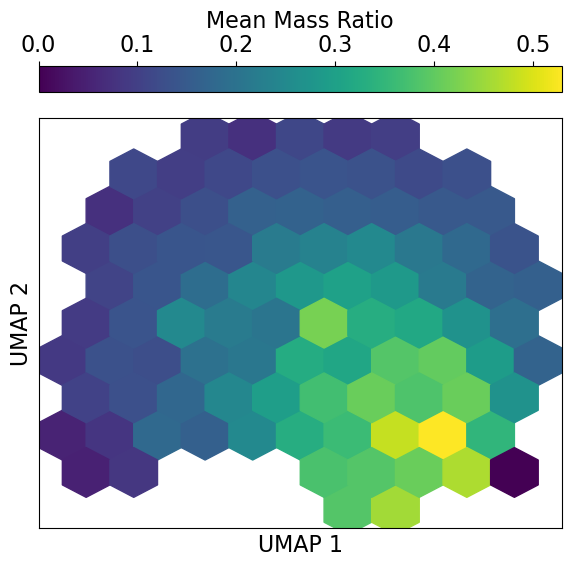}
  \caption{Mean Mass Ratio}
  \label{fig:MergerUMAP8}
\end{subfigure}
\begin{subfigure}{.33\textwidth}
  \centering
  \includegraphics[width=1.0\linewidth]{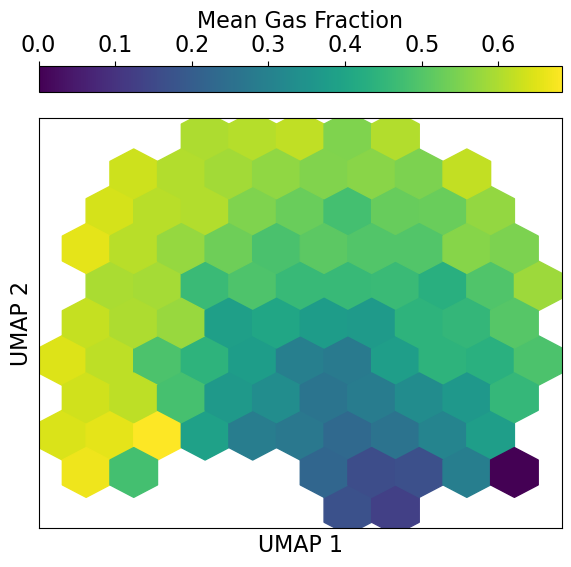}
  \caption{Mean Gas Fraction}
  \label{fig:MergerUMAP9}
\end{subfigure}%
\caption{Merger statistics using the UMAP representation; from left to right, mean edshift, mean mass ratio and  mean gas fraction. We recall the reader that only  bins with at least ten galaxies are plotted.}
\label{fig:MergerUMAP}
\end{figure*}

Figure~\ref{fig:MergerUMAP} examines aspects of the merger history using the UMAP representation as a guide. Although the merger catalogs are not explicitly used as input for the dimensionality reduction, the SFHs vectors contain information about mergers in the form of ex~situ mass. Figure~\ref{fig:MergerUMAP} shows the mean redshift of all the mergers a galaxy has undergone, the mean mass ratio of all the mergers (weighted by the mass of the progenitors) and the cold gas fraction (weighted by the maximum stellar mass of the merging objects). The three quantities show clear and smooth gradients, consistent with the ability of the dimensionality reduction to group galaxies with similar merger events. The bottom right corner is dominated by galaxies that have undergone recent mergers with a high mass ratio, consistent with the larger ex situ contribution seen in Fig \ref{UMAPSFH}. An interesting result is the anti-correlation with the mean cold gas fraction, which is larger at the left of the UMAP, a region populated by galaxies with small ex~situ component and fast star formation timescales (see Fig~\ref{fig:umapgen_d}). This could be consistent with intense cold gas accretion events at high redshift inducing starburst followed by a fast shut down of the SFR due to exhaustion of gas reservoir (see, e.g., \citealt{Ceverino2015, Zolotov2015}). However,  we recall that this value is normalized by the mass of the accreted systems, which are generally small for galaxies with non significant ex~situ contribution, so could be an artifact due to the chosen normalization.

\subsection{Kinematics and morphology in the UMAP plane}

\label{UMAP_cats}

Now that we have confirmed that the dimensionality-reduced UMAP plane captures relevant information from the input SFH catalog, in this section we  investigate whether it also reveals correlations with other fundamental galaxy properties—such as kinematics and morphology — to which the UMAP should, in principle, be agnostic.

\begin{figure*}
\centering
\begin{subfigure}{.245\textwidth}
  \centering
  \includegraphics[width=1.0\linewidth]{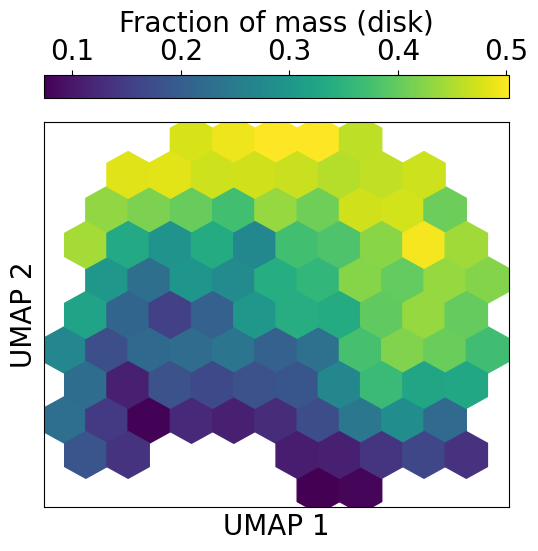}
  \caption{Disk contribution.}
  \label{fig:KineUMAP1}
\end{subfigure}%
\begin{subfigure}{.245\textwidth}
  \centering
  \includegraphics[width=1.0\linewidth]{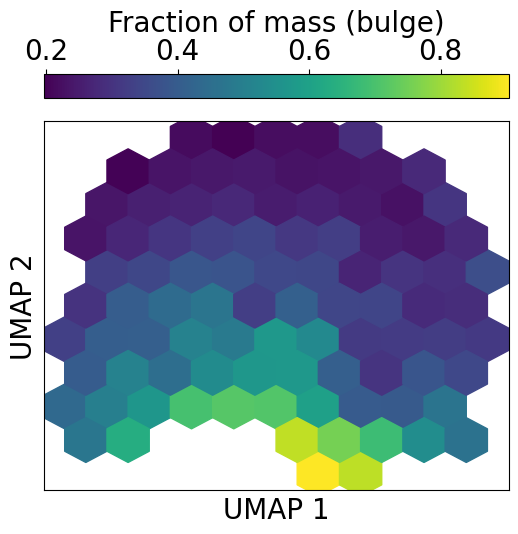}
  \caption{Bulge contribution.}
  \label{fig:KineUMAP2}
\end{subfigure}
\begin{subfigure}{.245\textwidth}
  \centering
  \includegraphics[width=1.0\linewidth]{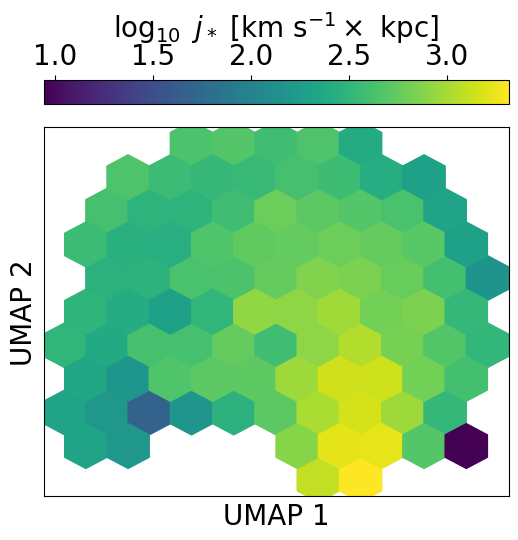}
  \caption{Specific Angular Momentum.}
  \label{fig:KineUMAP3}
\end{subfigure}
\begin{subfigure}{.245\textwidth}
  \centering
  \includegraphics[width=1.0\linewidth]{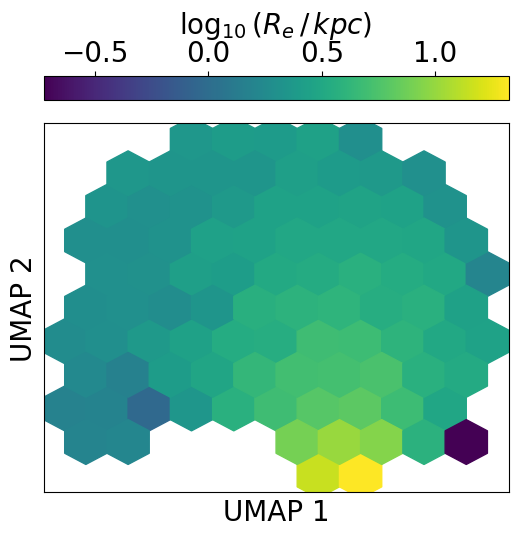}
  \caption{Stellar projected sizes ($R_e$).}
  \label{fig:SizeUMAP}
\end{subfigure}
\caption{Kinematic parameters and projected size using UMAP representation. Disk and bulge contribution to the stellar mass and specific angular momentum. Stellar sizes ($R_e$) are measured using the half-mass radii projected along the box z-axis. We only represent bins with at least ten galaxies.}
\label{fig:KineUMAP}
\end{figure*}

Figure~\ref{fig:KineUMAP} presents the UMAP plane color-coded by different kinematic parameters and projected sizes: the fraction of mass contained in a rotating disk (\ref{fig:KineUMAP1}), in a bulge (\ref{fig:KineUMAP2}) and the specific angular momentum (\ref{fig:KineUMAP3}). Disk-dominated galaxies are primarily located in the upper half of the UMAP plane, consistent with systems undergoing secular evolution and experiencing fewer interactions. In contrast,  galaxies with significant bulge components and fewer stars in rotating disks are predominantly located in the bottom-right corner of the UMAP plane. Interestingly, there is an important population of bulge-dominated galaxies with large angular momentum values (i.e., fast rotators), likely as a result of interactions with other galaxies. Although this finding may seem counter-intuitive, as one would typically expect rotation to be diminished by mergers, recent works have also shown that mergers can enhance disks and trigger star formation episodes due to significant gas accretion \citep{10.1093/mnras/stac2586}. Recent results regarding angular momentum and morphology can be seen in \cite{2023MNRAS.518.1002R}. Regarding sizes, Figure~\ref{fig:SizeUMAP} shows the representation of the half-mass radii ($R_{e}$) for each galaxy. The projection has been done along the z-axis with the intention of simulating a random projection.  The dependence with galaxy size in the UMAP plane is not as clear as with other quantities, but the largest galaxies are located in the same bottom right region where fast rotating massive galaxies are mostly found.
In Figures \ref{fig:KineUMAP3} and \ref{fig:SizeUMAP}, UMAP is successful in isolating very anomalous objects situated in the lower right corner, probably corresponding to recently formed objects in the simulation. We leave them in the plot for completeness.

\begin{figure*}
\begin{subfigure}{.245\textwidth}
  \centering
  \includegraphics[width=1.0\linewidth]{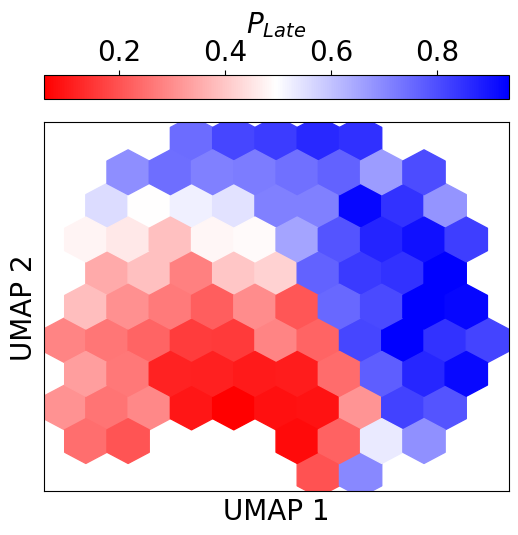}
  \caption{Probability of type $P_{\textrm{Late}}$.}
  \label{fig:MorphUMAP1}
\end{subfigure}
\begin{subfigure}{.245\textwidth}
  \centering
  \includegraphics[width=1.0\linewidth]{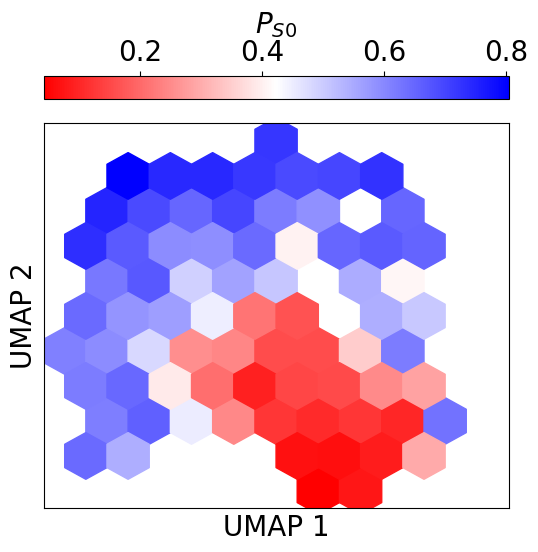}
  \caption{Probability of S0 for $P_{\textrm{Late}}<0.5$.}
  \label{fig:MorphUMAP2}
\end{subfigure}
\begin{subfigure}{.245\textwidth}
  \centering
  \includegraphics[width=1.0\linewidth]{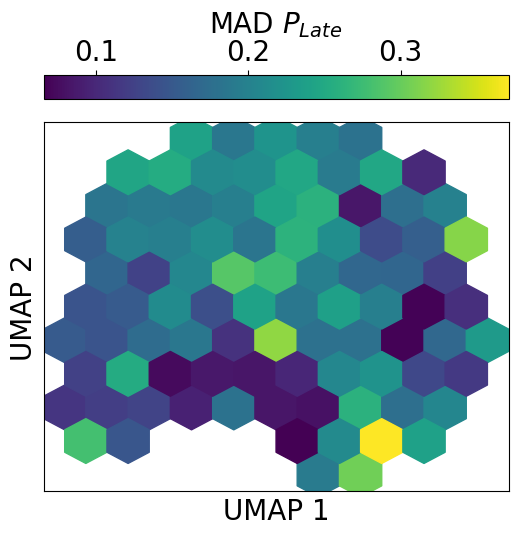}
  \caption{MAD of $P_{\textrm{Late}}$.}
  \label{fig:MorphUMAP1MAD}
\end{subfigure}
\begin{subfigure}{.245\textwidth}
  \centering
  \includegraphics[width=1.0\linewidth]{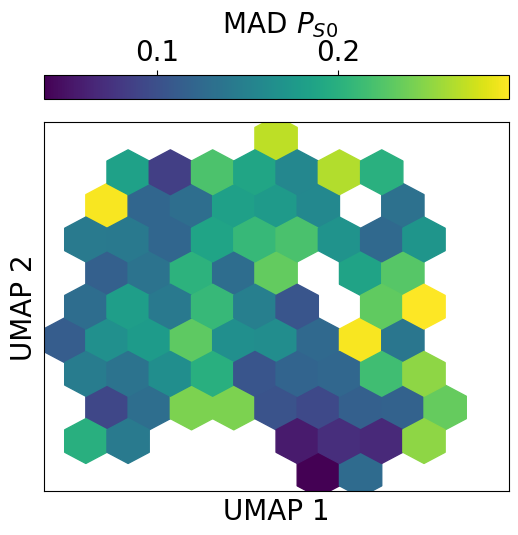}
  \caption{MAD of $P_{\textrm{S0}}$ for $P_{\textrm{Late}}<0.5$.}
  \label{fig:MorphUMAP2MAD}
\end{subfigure}
\caption{Morphology parameters, $P_{\textrm{Late}}$ and $P_{\textrm{S0}}$ using the UMAP representation along with dispersion in each bin using MAD (median absolute deviation) of both quantities. We only represent bins with at least ten galaxies.}
\label{fig:MorphUMAP}

\end{figure*}

Morphological parameters are closely related to kinematic properties. Figure~\ref{fig:MorphUMAP} shows the UMAP plane color-coded by morphological indicators. Figure~\ref{fig:MorphUMAP1} displays the probability of  a galaxy being late type ($P_{\text{Late}}$), with these systems predominantly located in the upper right part of the diagram. In contrast, early-type galaxies tend to occupy the lower and left regions. This spatial distribution aligns with expectations: late-type galaxies are typically disk-dominated, while early type galaxies are more bulge-dominated. Figure~\ref{fig:MorphUMAP2} shows the probability of a galaxy being S0 ($P_{\text{S0}}$), but only for systems filtered to have $P_{\text{Late}} < 0.5$, i.e., likely early-type galaxies. Interestingly, most elliptical (E) galaxies are found in the bottom-right corner, a region associated with high ex~situ mass contribution and large specific angular momentum. This suggests that the fast-rotating early-type galaxies in this region are predominantly elliptical rather than S0s. 

Figures \ref{fig:MorphUMAP1MAD} and \ref{fig:MorphUMAP2MAD} show the statistical dispersion of $P_{\text{Late}}$ and $P_{\text{S0}}$ in the UMAP plane, using the median absolute deviation (MAD). We choose MAD because it is more robust than the standard deviation. As expected, the dispersion is higher in regions with mixed populations, indicating greater morphological diversity within those bins.

\section{Clustering}
\label{Cluster}

Now that we have seen that dimensionality reduction helps to order similar galaxies together (in the UMAP parameter space), we apply clustering methods to understand the main formation channels of similar galaxies and to characterize their properties. Although several clustering methods have been explored, for simplicity, we show only the results of $k$-means here.

The clustering process can be approached in various ways, depending not only on the clustering algorithm itself  but also on the choice of input data and distance metrics. In this work, each galaxy is represented by a 200-dimensional vector (concatenation of 100 in~situ and 100 ex~situ components) and exhibits an inherent temporal structure, since the vectors reflect SFHs. 

Although $k$-means clustering typically relies on Euclidean distance, which is often suboptimal for time series or high-dimensional data, we opted for a direct clustering approach: applying $k$-means directly to the full 9648 × 200 input matrix. Despite the potential limitations of the Euclidean metric, this choice is justified by the normalization of the SFHs, which makes the Euclidean distance a reasonably effective measure of similarity in our context\footnote{Each galaxy is treated as a 200-dimensional vector, so the Euclidean distance used to measure similarity is calculated using the standard $L^2$-norm in $\mathbb{R}^{200}$.}. To assess the impact of dimensionality reduction, we also tested an intermediate approach: reducing the 200D input to 10D using UMAP, followed by $k$-means or hierarchical clustering. In particular, we find that this approach yields results comparable to those obtained without any dimensionality reduction. Although clustering on the 2D UMAP projection obtained in Section \ref{UMAP} could also be possible, it risks discarding relevant structural information present in the higher-dimensional data. Therefore, we reserve the final 2D UMAP projection solely for visualizing the clustering results. 

One of the important aspects of the $k$-means technique is to choose the number of clusters adequately. A very standard procedure is to create the ``elbow plot'', where the total distortion is plotted against the number of clusters. Another standard procedure is the so called ``average silhouette'' method \citep{ROUSSEEUW198753}. If we display the silhouette metric averaged over all observations against the number of clusters, the optimal $k$ should maximize the average silhouette. From the elbow and silhouette plots shown in the Appendix Figure \ref{fig:ElbowKmeans}, $k$ from 3 to 6 clusters seems to be an appropriate choice. Based on the subsequent morphological and kinematic analysis (see Section \ref{ClusterKDE}), we have chosen to set the number of clusters to 6.

\begin{figure*}
\begin{subfigure}{.33\textwidth}
  \centering
  \includegraphics[width=1.0\linewidth]{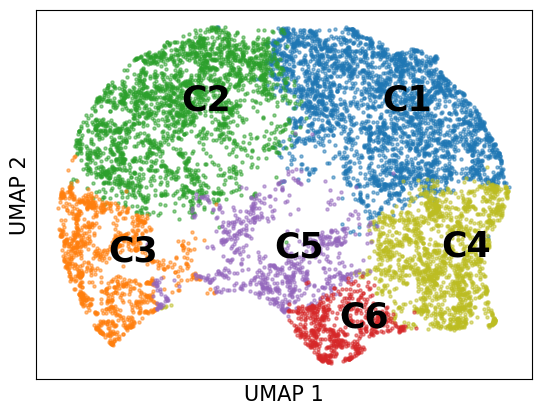}
  \caption{TNG100 and TNG50}
  \label{fig:ClusterKmeans1}
\end{subfigure}%
\begin{subfigure}{.33\textwidth}
  \centering
  \includegraphics[width=1.0\linewidth]{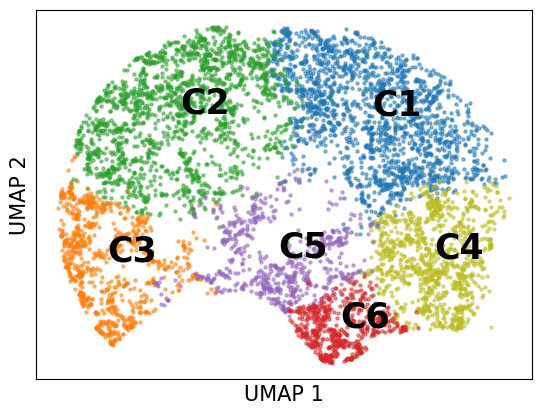}
  \caption{TNG100}
  \label{fig:ClusterKmeans2}
\end{subfigure}
\begin{subfigure}{.33\textwidth}
  \centering
  \includegraphics[width=1.0\linewidth]{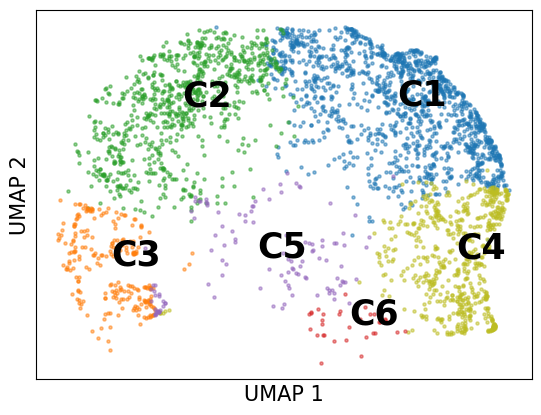}
  \caption{TNG50}
  \label{fig:ClusterKmeans3}
\end{subfigure}%
\caption{Results of $k$-means clustering with $k$ set to 6. The result is displayed in the UMAP 2D parametric space for the merged data TNG100 $\&$ TNG50 (left), TNG100 (middle) and TNG50 (right) separately.}
\label{fig:ClusterKmeans}
\end{figure*}

Figure \ref{fig:ClusterKmeans} shows the location of the galaxies that belong to the 6 different clusters in the 2D UMAP representation. There is little to no overlap of the clusters, meaning that the 2D UMAP is indeed a good representation of the higher-dimensional data  used to make the clustering. In the following subsections, we analyze the distinctive features of each cluster from different perspectives. In the Appendix Figure \ref{fig:ClusterWard} we show an analog of Figure \ref{fig:ClusterKmeans} but using hierarchical clustering with Euclidean distance and Ward linkage criterion. As we can see, the results are very similar, but there is a little more overlap between clusters in some regions.

\subsection{Star formation histories of the clusters}

\label{ClusterSFH}

\begin{figure*}
\centering
         \includegraphics[width=0.33\textwidth]{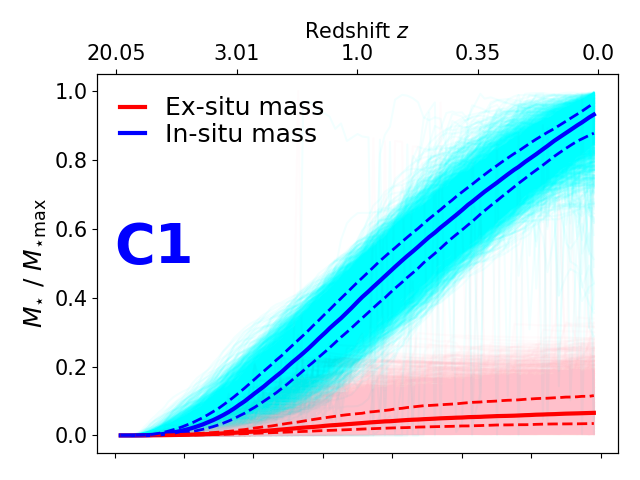}
         \includegraphics[width=0.33\textwidth]{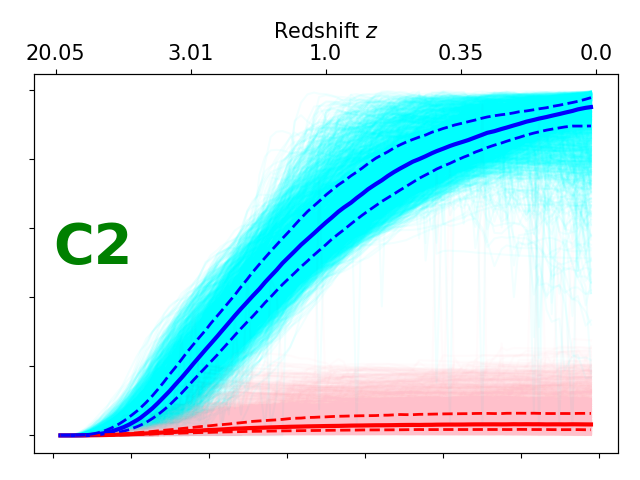}
         \includegraphics[width=0.33\textwidth]{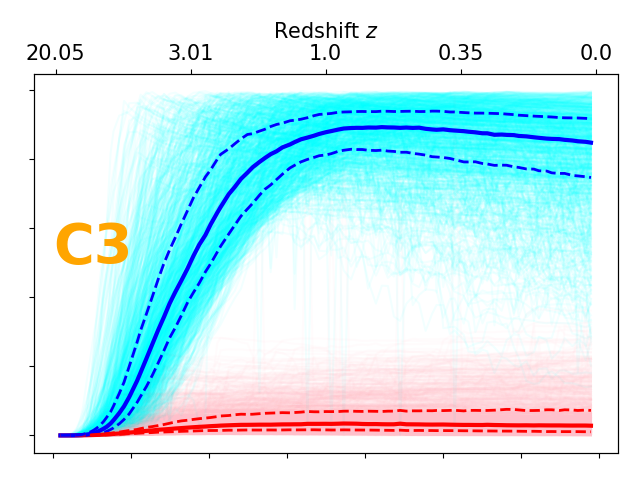}
         \includegraphics[width=0.33\textwidth]{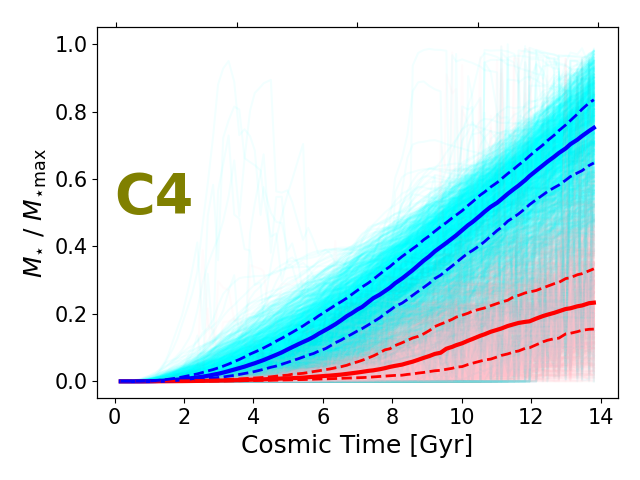}
         \includegraphics[width=0.33\textwidth]{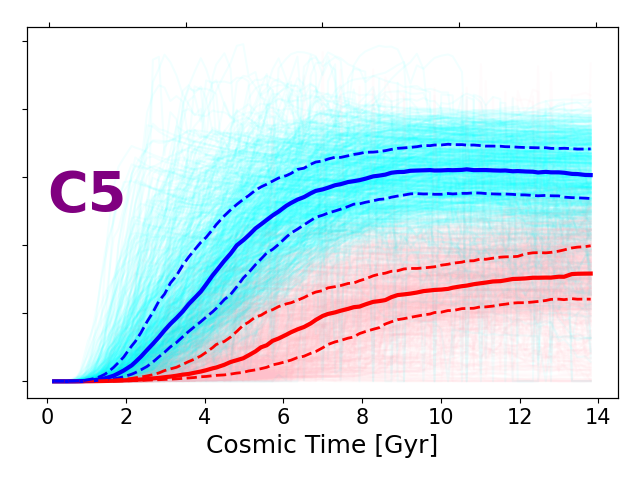}
         \includegraphics[width=0.33\textwidth]{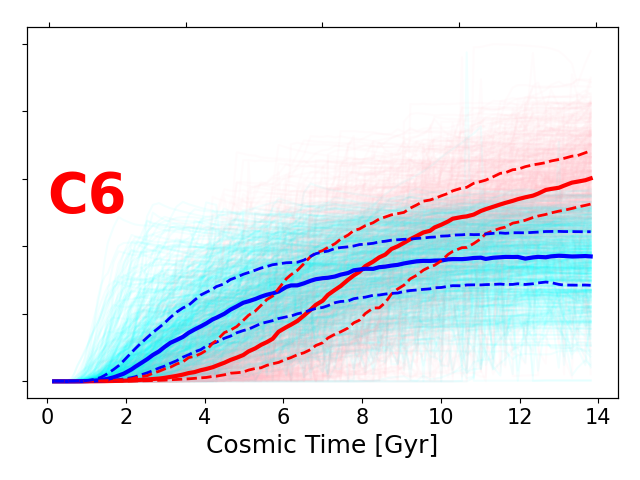}
        \caption{Star formation histories for galaxies belonging to the 6 clusters. In addition to the individual SFHs (light blue and pink for in~situ and ex~situ mass, respectively), we display the median mass (thick line) and the 25\% and 75\% percentiles (dashed lines).}
        \label{fig:SFHCluster}
\end{figure*}

Figure \ref{fig:SFHCluster} shows the SFHs curves of all galaxies belonging to each of the 6 clusters defined above. To summarize the information, we also plot the median and the 25\% and the 75\% percentiles. There are clear differences, but smooth variation on the SFH profiles from cluster 1 to cluster 6. In the following, we give a brief description of each cluster, and detailed information on their average properties is quoted in Table \ref{table:Mass}.

\begin{figure}
\centering
\includegraphics[width=1.0 \linewidth]{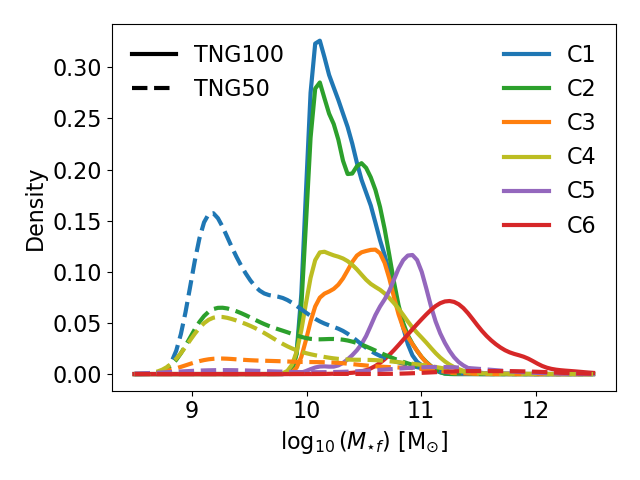}
\caption{\label{fig:MassKmeans} Kernel density estimate plots of final stellar mass (log scaled) under $k$-means clustering for clusters 1--6. Each curve represents a cluster and TNG100 and TNG50 are plotted separately. Area normalization under the curves has been made for all the curves combined, so that the sum of the areas of all the curves is one.}
\end{figure}

\begin{figure*}
    \begin{subfigure}{0.33\textwidth}
    \centering
    \includegraphics[width=1.0\linewidth]{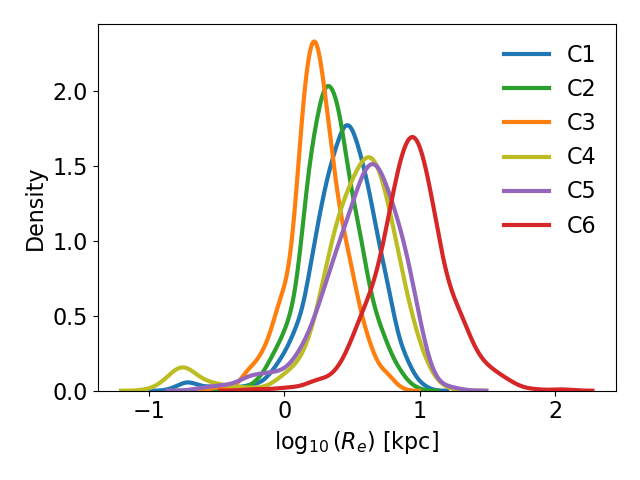}
         \caption{Half mass radius $R_e$ (log scaled).}
         \label{fig:SizeKDEKmeans}
    \end{subfigure}
    \begin{subfigure}{0.33\textwidth}
    \centering
     \includegraphics[width=1.0\linewidth]{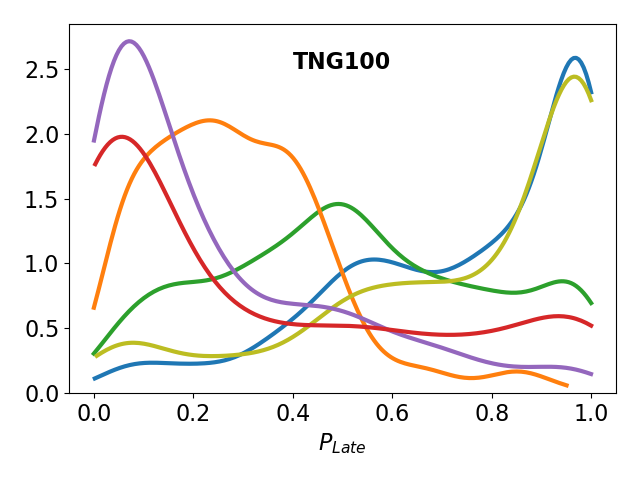}
     \caption{$P_{\textrm{Late}}$ for all galaxies.}
     \label{fig:MorphKmeans1}   
    \end{subfigure}
    \begin{subfigure}{0.33\textwidth}
    \centering
     \includegraphics[width=1.0\linewidth]{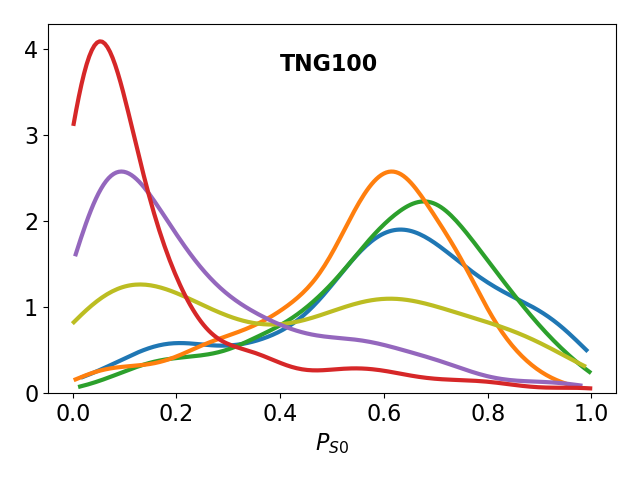}
     \caption{$P_{S0}$ for $P_{\textrm{Late}}<0.5$.}
     \label{fig:MorphKmeans2}   
    \end{subfigure}
    \begin{subfigure}{0.33\textwidth}
    \centering
     \includegraphics[width=1.0\linewidth]{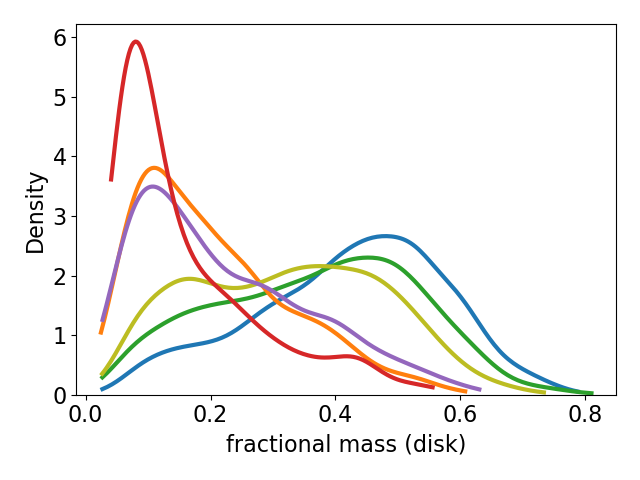}
        \caption{Disk contribution (fraction of mass).}
        \label{fig:KineKmeans1}
    \end{subfigure}
    \begin{subfigure}{0.33\textwidth}
    \centering
      \includegraphics[width=1.0\linewidth]{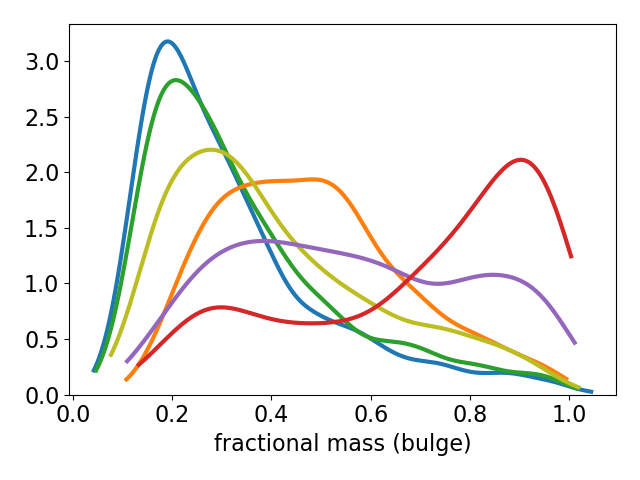}
        \caption{Bulge contribution (fraction of mass).}
        \label{fig:KineKmeans2}  
    \end{subfigure}
    \begin{subfigure}{0.33\textwidth}
    \centering
     \includegraphics[width=1.0\linewidth]{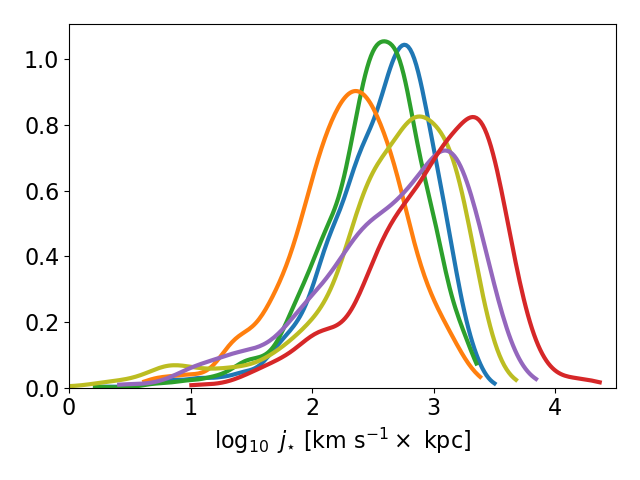}
        \caption{Specific angular momentum.}
        \label{fig:KineKmeans3}
    \end{subfigure}
     \caption{Kernel density estimate plots of different parameters under $k$-means clustering for clusters 1--6. The top left panel corresponds to the half mass radius $R_e$ (log scaled) density curve for each cluster. The top center and top right panels correspond to morphology (only TNG100 data) parameters $P_{\textrm{Late}}$ for all galaxies and $P_{S0}$ for early type galaxies ($P_{\textrm{Late}}<0.5$). The panels on the bottom display the fractional mass of stars associated to the disk (left) and the bulge (center) as well as the specific angular momentum (right) for clusters 1--6. For all plots each curve is normalized separately.}
    \label{fig:KDE}
\end{figure*}

\begin{table}
\caption{Number of galaxies, stellar mass and star formation rate of each cluster.}           
\label{table:Mass}      
\centering
\begin{tabular}{c c c c c}

\hline \hline

\multirow{2}{*}{TNG100 + TNG50} & \multicolumn{2}{c}{Galaxies} & Stellar Mass & SFR50\\
 & n & \%  & $[\log_{10}\mathrm{M}_{\odot}]$ & $[\mathrm{M}_{\odot}/yr]$ \\
 
\hline  

Cluster 1 & 3158 & 33 & 10.12 & 0.61 \\
Cluster 2 & 2500 & 26 & 10.23 & 0.11 \\
Cluster 3 & 1068 & 11 & 10.42 & 0.00 \\
Cluster 4 & 1573 & 16 & 10.23 & 0.85 \\
Cluster 5 &  748 &  8 & 10.86 & 0.00 \\
Cluster 6 &  601 &  6 & 11.28 & 0.07 \\
Total/Average & 9648 & 100 & 10.28 & 0.23 \\

\hline\hline 
       
\multirow{2}{*}{TNG100} & \multicolumn{2}{c}{Galaxies} & Stellar Mass & SFR50\\
 & n & \%  & $[\log_{10}\mathrm{M}_{\odot}]$ & $[\mathrm{M}_{\odot}/yr]$ \\
 
\hline  

Cluster 1 & 1789 &  28 &  10.28 & 1.10  \\
Cluster 2 & 1706 &  26 &  10.32 & 0.18 \\
Cluster 3 &  827 &  13 &  10.48 & 0.00  \\
Cluster 4 &  964 &  15 &  10.40 & 1.30  \\
Cluster 5 &  628 &  10 &  10.85 & 0.00  \\
Cluster 6 &  567 &   9 & 11.27 & 0.04  \\
Total/Average     & 6481 & 100 &  10.44 & 0.33  \\
                
\hline \hline 

\multirow{2}{*}{TNG50} & \multicolumn{2}{c}{Galaxies} & Stellar Mass & SFR50\\
 & n & \%  & $[\log_{10}\mathrm{M}_{\odot}]$ & $[\mathrm{M}_{\odot}/yr]$ \\
 
\hline  

Cluster 1 & 1369 & 43 & 9.45 & 0.24 \\
Cluster 2 &  794 & 25 &  9.61 & 0.08 \\
Cluster 3 &  241 &  8 &   9.76 & 0.00 \\
Cluster 4 &  609 & 20 & 9.51 & 0.41 \\
Cluster 5 &  120 &  4 &   10.92 & 0.00 \\
Cluster 6 &   34 &  1 &  11.56 & 3.33 \\
Total/Average & 3167 & 100 & 9.55 & 0.19 \\

\hline 

\end{tabular}

\tablefoot{Number of galaxies, median final stellar mass and star formation rate for each cluster, separated into TNG100 and TNG50 galaxies, as well as the combined sample. We warn the reader that the statistics of the combined sample should be interpreted with caution due to the different mass distribution of each simulation.}   

\end{table}

\begin{itemize}
    
    \item Cluster 1 (C1): This is the most populated cluster, containing 3158 galaxies (33\% of the total sample). Galaxies in this cluster are characterized by a secular evolution. The main mass contribution comes from constant in~situ star formation, which is still taking place, while there is very little ex~situ mass accretion. 
    
    \item Cluster 2 (C2): Including 2500 galaxies (26\%). It can be seen as a transition cluster from 1 (constant SFR) to cluster 3 (quenching process). These galaxies are still dominated by in~situ mass and are on their way to quenching.

    \item Cluster 3 (C3): 1068 galaxies (11\%). These are galaxies that have ceased star formation a long time ago. The in~situ mass peaked around 6-8 Gyr from the present time. The ex~situ contribution is  limited, but some  have suffered some mass loss.

    \item Cluster 4 (C4): 1573 galaxies (16\%). Galaxies in this cluster are forming stars, but the ex~situ mass contribution is relevant. It could be interpreted as a transition between  cluster 1 (dominated by secular evolution) and clusters 5-6 (dominated by major mergers).
    
    \item Cluster 5 (C5): 748 galaxies (8\%). It is similar to cluster 3 (quenched galaxies) but with a significant ex~situ contribution, relatively constant in time.
    
    \item Cluster 6 (C6): the least populated cluster with only 601 galaxies (6\%), where most major merger events take place, resulting in a mass assembly history dominated by  ex~situ contribution, especially at later times. 
\end{itemize}

Figure~\ref{fig:SFHCluster100} shows the same clustering exercise but using only TNG100 data. The similarity between the SFHs of the clusters with and without including TNG50 provides additional evidence in favor of merging both datasets: adding TNG50 does not alter the data structure, but rather contributes with statistics and diversity in the lower mass range.

Due to the ambiguity in deciding the number of clusters (as shown in Figure~\ref{fig:ElbowKmeans}), we calculated the average SFHs with both $k=5$ and $k=7$ clusters. If we fix $k=5$, the C5 and C6 clusters merge in a single one. We find that keeping the two populations separated is  physically interesting because C5 has a significant ex situ contribution, while C6 is dominated by the ex situ contribution, and this difference is important. When setting $k=7$, the original C3 population is divided into two groups, one very similar to the original C3 and the new cluster showing more recent quenching and less mass loss. Although this subdivision provides additional detail, we find that it does not reveal a qualitatively new evolutionary pathway that would justify the increased complexity of the following analysis. Therefore, we believe that $k=6$ is an optimal trade-off between complexity and diversity.

Although the six clusters identified in this work do not represent unknown galaxy populations, it is remarkable that these groups emerge naturally from a fully unsupervised analysis based solely on the SFHs of galaxies. The resulting populations closely resemble well-known observational classes, including star-forming disk galaxies, transition systems, quenched S0s, massive ellipticals, and merger-dominated systems, which have been extensively identified through morphology, kinematics, and scaling relations \citep[e.g.,][]{Blanton2009, Cappellari2016, DS2018, DS2022}. This suggests that the diversity of present-day galaxy populations may largely arise from a limited number of underlying assembly histories encoded in their SFHs. Interestingly, while the individual SFHs exhibit complex temporal variations, including burst-like episodes of enhanced star formation, the median SFHs of the different clusters do not show pronounced peaks, suggesting that, although individual galaxies may experience episodes of intense star formation, these events do not occur at a common preferred epoch across the galaxy populations.

\subsection{Dependence of galaxy properties on clustering}

\label{ClusterKDE}

To shed some light on the main differences between the properties of galaxies belonging to different clusters, we analyze their normalized distributions. This is shown in Figures \ref{fig:MassKmeans} and \ref{fig:KDE}, where we display kernel density estimate (KDE) for different physical variables and how they are distributed among clusters 1--6. We start our analysis with  physical variables taken from the SFHs catalog. Figure~\ref{fig:MassKmeans} shows the mass distribution for the 6 groups, where there are clear differences between them. The differences in volume and mass resolution between TNG50 and TNG100 have a strong impact on the mass distribution of galaxies. Therefore, we plot separately the mass distribution for each simulation set. C1, C2 present the lowest masses, C3 and C4 have intermediate mass distributions, and C5 and C6, where the ex~situ contribution is more important,  show the highest mass values. We also note that most of the galaxies belonging to C5 and C6 come from the TNG100 run ($\sim$ 10\% in both), while they are very scarce in TNG50 ($\sim$4\% and  $\sim$1\%, respectively) due to its smaller volume. We remind the reader that the input for the clustering analysis 
are the SFHs normalized by mass, meaning that the evident correlation between clusters and stellar mass is fully driven by differences in mass assembly histories.

In the following plots in Figure \ref{fig:KDE}, we study how the morphological and kinematic properties - which were not used as input information for the clustering - vary among the groups found by $k$-means. The analysis is also performed using the KDE density plot, where each curve is normalized individually. For the morphological classification, we only use the TNG100 dataset, as the morphological catalog is not available for TNG50 galaxies at $z=0$.

Figure~\ref{fig:SizeKDEKmeans} displays additional information that can be compared with Figure~\ref{fig:MassKmeans} as massive galaxies tend to be larger in size. Galaxies in C1, C2, and C3  are the most compact galaxies and correspond to small to medium masses. C4 and C5 have intermediate radii, but C5 galaxies are significantly more massive than those present in C4. To conclude, galaxies in C6 are large both in mass and in size. Figures \ref{fig:MorphKmeans1} and \ref{fig:MorphKmeans2} show the morphological classification distribution for the different groups. Regarding the separation between ETGs and LTGs (Figure~\ref{fig:MorphKmeans1}), C1 and C4 are clearly dominated by LTG galaxies, while C3, C5, and C6 show mainly ETG morphologies. C2 shows a very wide distribution, in agreement with the fact that they are galaxies on their way to quenching. In Figure~\ref{fig:MorphKmeans2}, we  show the distribution of $P_{S0}$ - the probability of being S0 rather than pure E - for galaxies that have $P_{\text{Late}}<0.5$. Lenticular galaxies are more common in C1, C2, and C3, while C5 and C6 are dominated by E galaxies. C4 shows a very wide distribution, but note that most galaxies in this cluster have $P_{\text{Late}}> 0.5$ and therefore are not represented in this plot.

To end this section, we perform the same cluster analysis for the kinematic variables in Figures \ref{fig:KineKmeans1}, \ref{fig:KineKmeans2}, and \ref{fig:KineKmeans3}. In particular, Figure \ref{fig:KineKmeans1} displays the fractional mass in the disk, while Figure \ref{fig:KineKmeans2} shows the fractional mass in the bulge. C1, C2, and C4 are clearly disk dominated, while C6 is clearly bulge dominated, consistent with the morphological classifications presented in Figures \ref{fig:MorphKmeans1} and \ref{fig:MorphKmeans2}. C5 shows a more mixed contribution of stars, consistent with distortions associated with mass accretion and interactions. Finally, C3 has a very small contribution to the disk, but it is not dominated by the bulge component either, consistent with the S0 classification. The figure on the right displays the specific angular momentum. The difference in this quantity within different clusters is not as evident as in other properties. C6 tends to have the fastest rotators on average, but its angular momentum distribution shows a large dispersion compared to the other clusters. This is somehow unexpected since C6 contains the most massive and merger-dominated systems, and they are expected to be slow rotators, but works like \cite{10.1093/mnras/stac2586} show that the opposite can also happen. Thus, the two effects are consistent with the high dispersion present in this cluster. The distribution of the angular momentum of the galaxies in C3 also has a large scatter and tends to have the slowest rotators. In contrast, the rest of the clusters tend to have less dispersed angular momentum distributions and relatively moderate levels of rotation.

\subsection{Scaling relations of the clusters}

\label{ClusterScaling}

\begin{figure*}
    \centering
    \includegraphics[width=\textwidth]{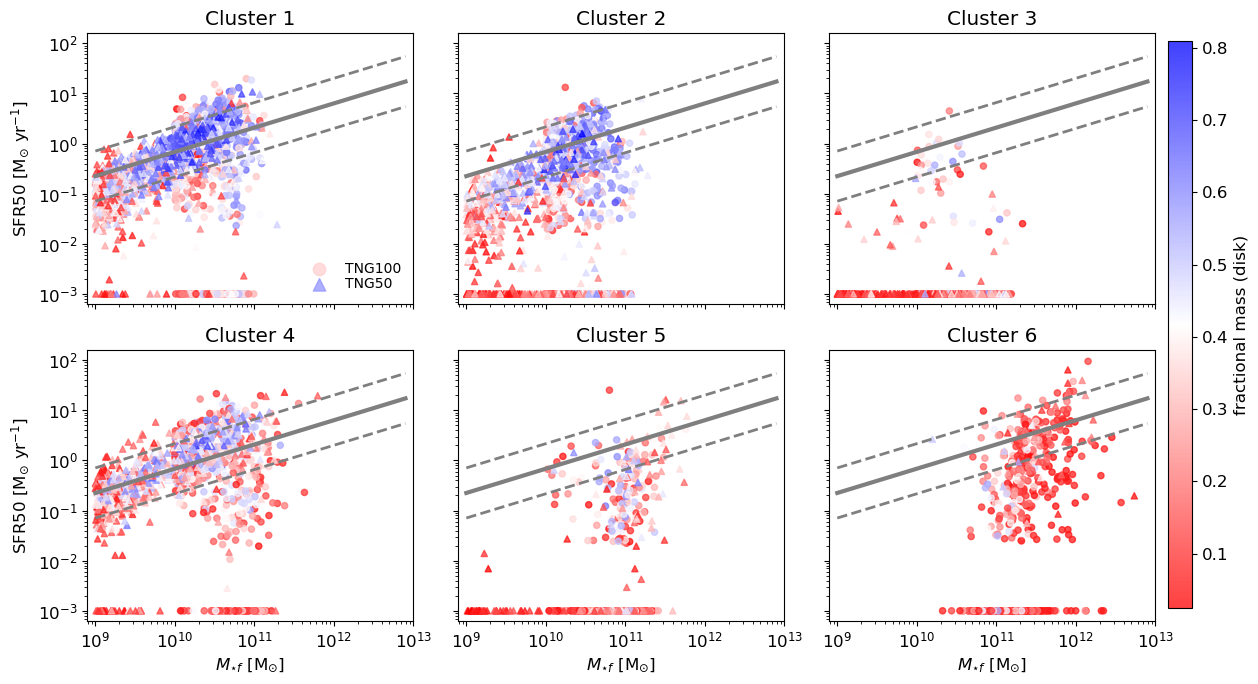} 
        \caption{SFR-Mass relation for the six galaxy clusters (one for each panel). As a reference, the MS of \cite{2014ApJS..214...15S} is over-plotted (gray line), with 0.5 dex scatter (dashed line). All galaxies with SFR less than $10^{-3}$ $\mathrm{M}_{\odot}$ yr$^{-1}$ have been set to $10^{-3}$ for visualization purposes (many galaxies have 0 SFR). Galaxies are color coded according to  their fractional mass in disk  (see Section \ref{UMAP_cats} and Figure \ref{fig:KineUMAP}). Triangles show TNG50 galaxies while circles correspond to TNG100. }
        \label{fig:SFR}
\end{figure*}

In this section, we analyze the location of the six clusters defined above in important scaling relations, such as the SFR--Stellar Mass plane or the Stellar Mass--Size relation. Figure \ref{fig:SFR} shows the SFR-Mass plane for the different clusters, color-coded by the fraction of mass contained in the disk. As a reference, the Main Sequence (MS) for galaxy formation from \cite{2014ApJS..214...15S} is also shown\footnote{$\log_{10}SFR(M_{\star},t) = (0.84-0.026t) \log_{10}M_{\star}-(6.51- 0.11t)$}. For visualization purposes, we set to $10^{-3}$ the SFR of all galaxies with less than $10^{-3}$ $\mathrm{M}_{\odot}$ yr$^{-1}$. This is because there are a significant number of galaxies with SFR equal to zero, in particular, in clusters where star formation is quenched. The mass differences from TNG50 and TNG100 become evident as a vertical structure around 10$^{10}$ $\mathrm{M}_{\odot}$.  We also summarize the fraction of galaxies above, below, and on top of the MS for each cluster in Table \ref{table:TNG100}. Most of the galaxies from C1 and C4 are located around the MS (> 70\% of them, see Table \ref{table:TNG100}), while cluster 2 shows galaxies all over the SFR-mass plane. In contrast, most of the galaxies from C3 and C5 are located below the MS (> 80\%), in agreement with their quenched SFHs. Finally, C6 includes  most of the massive galaxies in the sample and shows a diverse population, suggesting that mergers can both enhance or quench star formation. There is a clear correlation between the fraction of mass contained in the disk and the location on the SFR-Mass plane, with most galaxies below the MS showing no disk component. This is in agreement with previous studies suggesting a strong correlation between structural transformations and quenching mechanisms, see \cite{Bluck2014, DS2022, Kauffmann2003a, Kauffmann2003b, Kauffmann2003c, Kormendy_2009, Tacchella2015}. Interestingly, galaxies belonging to C6 are bulge dominated, regardless of their location in the SFR-Mass plane. This could be explained if mergers disrupt the disk structure before the quenching of star formation takes place.

\begin{figure*}
         \includegraphics[width=\textwidth]{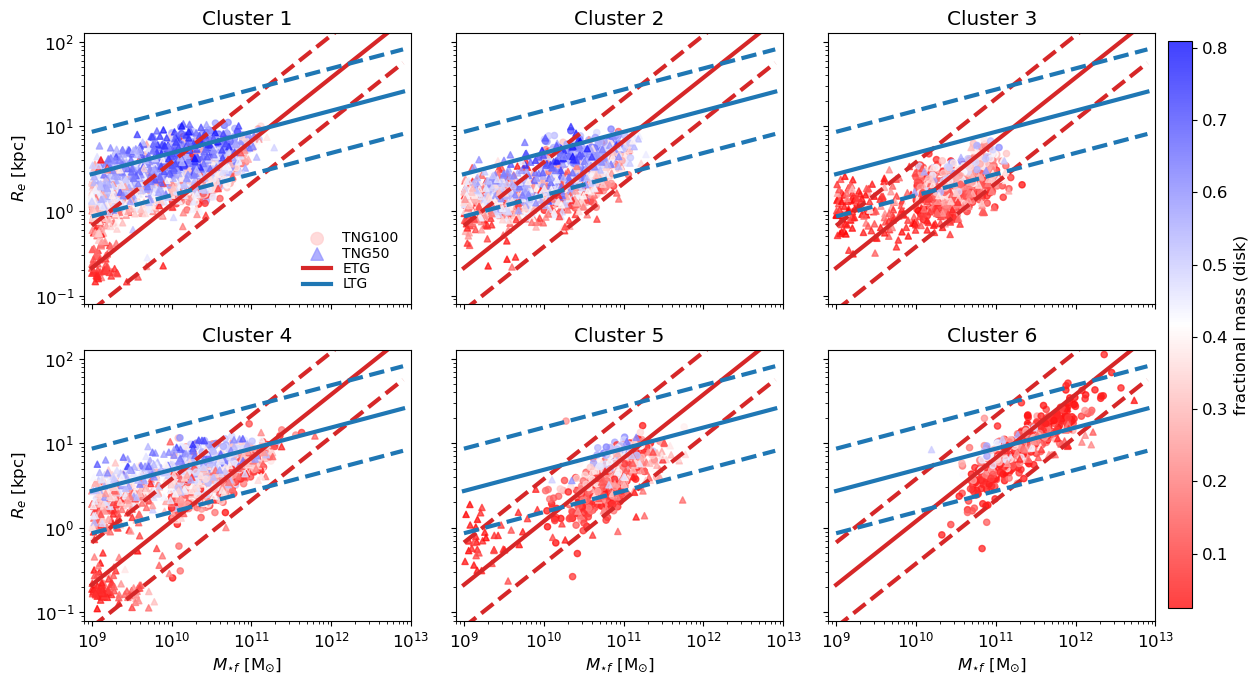}
        \caption{$R_e$-Mass scatterplot. $R_e$ is calculated as half mass radius. The lines correspond to the mass-size relations for early (red line) and late type (blue line) galaxies taken from \cite{2014ApJ...788...28V}. The relations used are $ r(m_{\star}) \mathrm{/kpc}  = A \cdot m^{\alpha}_{\star}$ where $m_{\star}=M_{\star} /7 \times 10^{10} \, \mathrm{M}_{\odot}$. We use the values provided for $z=0.25$, in particular the parameters $\log A=0.6$ and $\alpha=0.75$ for early-type galaxies and $\log A=0.86$ and $\alpha=0.25$ for late-type galaxies. Errors (dotted lines) are 0.5 dex. Each cluster is shown in a different figure and the color code corresponds to the fractional mass in disk studied in Section \ref{UMAP_cats} and Figure \ref{fig:KineUMAP}. Triangles show TNG50 galaxies while circles correspond to TNG100.}
        \label{fig:SizeKmeans}
\end{figure*}

Figure \ref{fig:SizeKmeans} shows the Mass-Size plane for the 6 clusters. We show the effective radius ($R_e$) as the half mass radius. The color code corresponds to the fraction of mass contained in the disk. We also add analytic Mass--Size relations for $z=0.25$ (the closest available to our simulated data at $z=0$) from \cite{2014ApJ...788...28V}. The Mass--Size relations are clearly different for bulge-dominated, early-type (ETG), and disk-dominated, late-type galaxies (LTG), in agreement with previous results from the literature \citep{Shen2003,2014ApJ...788...28V}. Most of the C1 galaxies  follow the LTG mass-size relation and are dominated by the disk component, as expected for star-forming galaxies. On the other hand, C3, C5, and C6 galaxies mostly follow the mass-size relation from ETGs and are dominated by the bulge component. This is consistent with the mass-size relation being driven by the SFH rather than the merging history (note that galaxies in these clusters are quenched but with different levels of ex~situ contribution). On the other hand, C2 and C4 galaxies show a mixture in their mass-size relation. Galaxies in C2 are on their way to quench, and their location in the mass-size plane is consistent with their structural composition (disk galaxies closer to the LTG size-mass relation, spheroidals being more compact). C4 is the most complex cluster in this relation, showing some galaxies on top of the LTG relation, as well as some galaxies on the ETG relation. Interestingly, this is the only cluster where a significant fraction of bulge-dominated galaxies follow the LTG relation, indicating that the ex~situ contribution (the main difference between C1 and C2) induces the presence of a bulge before affecting the mass-size relation. Another important point is the fact that most galaxies follow the observed mass-size relation, despite the contribution of the accreted ex~situ mass. This is especially remarkable for C6 galaxies, where the ex~situ contribution dominates the SFHs, but closely follows the observed mass-size relation.  This may be an indicator that a significant fraction of galaxies used to calibrate this relation have indeed undergone relevant merging episodes.

\begin{table*}
\caption{Galaxy clusters in TNG100 and TNG50 separately.}         
\label{table:TNG100}      
\centering
\begin{tabular}{c c c c c c c c c c}

\hline \hline

\multirow{2}{*}{TNG100 + TNG50} & \multicolumn{3}{c}{Main Sequence} & \multicolumn{3}{c}{Above MS} & \multicolumn{3 }{c}{Below MS} \\ 
 & n & \%  Cluster & \% Total & n & \%  Cluster & \% Total & n & \%  Cluster & \% Total \\

 \hline

Cluster 1 & 2432 & 77 & 25  & 155 & 5 & 2 & 571 &  18 & 6\\
Cluster 2 & 1065 & 43 & 11  &  24 & 1 & 0.2 & 1411 & 56 & 15\\
Cluster 3 &   21 & 2 & 0.2  &   1 & 0.1 & 0.01 & 1046 &  98 & 11\\
Cluster 4 & 1100 & 70 & 11  & 192 & 12 & 2 & 281 &  18 & 3 \\
Cluster 5 &   59 & 8 & 0.6   & 4   & 0.5 & 0.04 & 685 & 92 & 7 \\
Cluster 6 &  126 & 21 & 1 & 17  & 3 & 0.2 & 458 & 76 & 5 \\
Total & 4803 & - & 50 & 68 & - & 4 & 4452 & - & 46\\

\hline\hline

\multirow{2}{*}{TNG100} & \multicolumn{3}{c}{Main Sequence} & \multicolumn{3}{c}{Above MS} & \multicolumn{3 }{c}{Below MS} \\ 
 & n & \%  Cluster & \% Total & n & \%  Cluster & \% Total & n & \%  Cluster & \% Total \\
 
\hline  

Cluster 1 &  1338 &75 & 21  & 141 & 8 & 2 &  310 &17 & 5  \\
Cluster 2 &  762 &45 & 12 &  23 & 1 & 0.4 &  921 & 54  &  14  \\
Cluster 3 &   21 & 3 & 0.3 &   1 & 0 & 0 &  805 & 97  & 12  \\
Cluster 4 &  615  &64 & 10 & 148 &15 & 2  &  201 & 21  & 3  \\
Cluster 5 &  38 & 6 & 0.6  &   1 & 0 & 0  &  589 & 94 & 9  \\
Cluster 6 &  107 &19 & 2  &  11 & 2 & 0.2  &  449 & 79 & 7   \\
Total    & 2881 & -- & 44  & 325 & -- & 5  & 3275 & -- & 51 \\
                
\hline \hline

\multirow{2}{*}{TNG50} & \multicolumn{3}{c}{Main Sequence} & \multicolumn{3}{c}{Above MS} & \multicolumn{3 }{c}{Below MS} \\ 
 & n & \%  Cluster & \% Total & n & \%  Cluster & \% Total & n & \%  Cluster & \% Total \\

\hline

Cluster 1 & 1094 & 80 & 35 & 14 & 1 & 0.4 & 261 &  19 & 8\\
Cluster 2 &  303 &  38 & 10 & 1 & 0.1 & 0 & 490 & 62 & 15\\
Cluster 3 &  0   &  0 & 0 &  0 & 0 & 0 & 241 &  100 & 8\\
Cluster 4 &  485 & 80 &  15 & 44 & 7 & 1 & 80 &  13 & 3 \\
Cluster 5 &   21 & 17 &  1 & 3 & 3 & 0.1 & 96 &  80 & 3 \\
Cluster 6 &   19 &  56 & 0.6  & 6 & 18 & 0.2 &  9 &  27 & 0.3 \\
Total & 1922 & - & 61 & 68 & - & 2 & 1177 & - & 37\\
                                
\hline
  
\end{tabular}

\tablefoot{For each cluster we calculate the number and fraction of galaxies in the Main Sequence (MS), below the MS and above the MS. Fractions are given as a percentage over the total number of galaxies in the dataset (TNG100 and TNG50 separately) and over the number of galaxies in each cluster for each dataset. For definition of the MS see Section \ref{ClusterScaling} and Figure \ref{fig:SFR}.}
\end{table*}

\section{Summary and conclusions}
\label{Conclusions}

In this work, we have applied unsupervised machine learning techniques --- dimensionality reduction and clustering --- to the star formation histories (SFHs) of $\sim$10000 galaxies drawn from the TNG50 and TNG100 cosmological simulations, with the goal of identifying the main galaxy formation channels in a data-driven way.

We have shown that UMAP is able to organize galaxies along their SFHs into a coherent two-dimensional representation, where position in the UMAP plane correlates strongly with physical galaxy properties that were used as input (e.g.,  formation timescales $t_{50}$ and $t_{90}$,  current star formation rate, merger history), but also with properties that were not (such as stellar mass, morphology, and kinematics). This demonstrates that the diversity of galaxy assembly histories can be encoded in a low-dimensional way - and that UMAP is an effective tool for visualizing it - and that the SFH leaves imprints on observational galaxy properties.

Applying $k$-means clustering to the SFH vectors, we identify six groups of galaxies with distinct assembly histories, characterized by different levels of star formation activity and ex~situ stellar mass fraction. These groups are well separated in the UMAP plane and show systematic differences in their physical properties, which can be summarized as follows:

\begin{itemize}

    \item \textbf{C1} comprises the largest galaxy population, with low stellar masses and a continuous in~situ SFR. These are disk-dominated, late-type galaxies located on or near the star-forming main sequence (~80\% of the galaxies in this group), with intermediate angular momentum and negligible ex~situ contribution. Their position in the size-mass plane is closely correlated with their fractional disk mass.

    \item \textbf{C2} represents a transition population between active star-forming and quiescent systems. Galaxies in this cluster show a declining SFR and morphologies ranging from late-type to S0, with a significant disk component, intermediate angular momentum and small ex~situ fraction. These galaxies occupy wide regions of the SFR-mass and size-mass planes.

    \item \textbf{C3} contains galaxies that quenched their star formation several Gyrs ago, with in~situ mass peaking at $\sim$6--8~Gyr lookback time and minimal ex~situ contribution. These are intermediate-mass ETGs (including a significant fraction of S0s) with low angular momentum, predominantly located below the main sequence (> 97\%) and in the ETGs loci of the mass-size relation.

    \item \textbf{C4} is similar to C1 in terms of ongoing star formation and late-type morphology, but with a more significant ex~situ mass contribution and higher angular momentum. The fraction of galaxies above and on the MS is similar to that of C1 ($\sim 80\% $) but some of the galaxies on the MS and on the loci of the LTGs size-mass relation   show relatively low values of fractional disk mass. This suggests that moderate accretion can induce bulge growth before affecting SFR or size.

    \item \textbf{C5} galaxies are quenched systems with a significant and sustained ex~situ mass fraction. They have large stellar masses and tend to show elliptical morphologies with relatively high angular momentum. They are mainly located below the MS (> 80\%) and follow the ETG mass-size relation. Compared to C3 galaxies, they show slightly larger masses, sizes, and SFRs, consistent with having undergone multiple minor mergers over their lifetimes.

    \item \textbf{C6} is the least populated but most extreme cluster, dominated by galaxies whose mass assembly is largely driven by ex~situ accretion through major merger events. These are the most massive systems in the sample, with elliptical morphologies, negligible disk component, and - counterintuitively -  the highest specific angular momentum values, consistent with a population of fast-rotating early-type galaxies.

\end{itemize}

A direct comparison between clusters with similar star formation activity but different ex~situ mass fractions provides further insight into the role of accretion-driven evolution. In the star-forming regime, C1 and C4 galaxies share similar positions along the main sequence and the late-type mass–size relation, yet differ in their assembly histories: while C1 systems are dominated by in~situ growth, C4 galaxies exhibit a non-negligible ex~situ contribution. This difference is reflected in their structural properties, with C4 galaxies showing lower disk fractions and higher angular momentum, suggesting that moderate accretion can induce morphological transformation without significantly altering their global star-forming properties. 

A similar trend is observed in the quenched population when comparing C3 and C5. Both clusters are dominated by galaxies below the main sequence and lying on the early-type mass–size relation, but C5 systems, with higher ex~situ mass fractions, tend to have larger stellar masses, sizes, and residual star formation. This indicates that cumulative minor mergers can drive continued mass growth and structural evolution after quenching, in contrast to the more passive evolution of C3 galaxies, whose assembly is largely in~situ.

Finally, C6 represents the extreme end of this sequence, where galaxy assembly is dominated by major mergers. These systems occupy the high-mass tail of the population and show properties consistent with fast-rotating ellipticals, highlighting the key role of major mergers in shaping the most massive galaxies and driving the transition to dispersion-dominated systems.

Overall, our analysis demonstrates that the diversity of galaxy formation histories in TNG can be organized into a small number of characteristic evolutionary pathways and that unsupervised learning provides a powerful and interpretable framework for linking galaxy assembly histories to their present-day observable properties. We leave for future work the analysis of these clusters with other properties like environment or velocity dispersion. 

We note that this methodology is very useful for investigating particular types of galaxies, as done in a companion paper (Granados-Buey et al. in prep) to find massive relic galaxies, galaxies that have remained frozen in time without significant accretion since their early formation at $z\sim 2$ \citep{Ferre-Mateu2017}.

\begin{acknowledgements}

CAF acknowledges financial support by Santalucía Chair of Analytics for Education. HDS acknowledges financial support by RyC2022-030469-I grant, funded by MCIU/AEI/10.13039/501100011033, and FSE+ and the Spanish Ministry of Science. JV-F acknowledges the financial support from the European Union - NextGenerationEU and the Spanish Ministry of Science and Innovation through the Recovery and Resilience Facility project ICTS-MRR-2021-03-CEFCA and the support from the grant PID2022-138896NA-C54. AFM acknowledges support from RYC2021-031099-I and PID2024-162088NB-I00 of MICIN/AEI. DSR acknowledges financial support from the Ministerio de Ciencia e Innovación (Spain) via Project grant CNS2024-154550. 
     
\end{acknowledgements}


\bibliographystyle{aa}
\bibliography{refs}

\begin{appendix}

\section{Data consistency}

\begin{figure}
    \centering
    \includegraphics[width=1.00\linewidth]{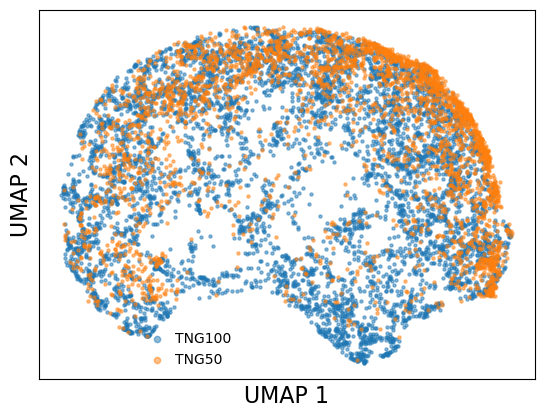}
    \caption{UMAP representation of both TNG100 and TNG50 data where the UMAP projection has been trained using only TNG100 data. }
    \label{fig:Umapmixed}
\end{figure}

Figure~\ref{fig:Umapmixed} shows the representation of TNG100 and TNG50 in a 2D projection fitted only using TNG100 data. The overlapping of both datasets and the similarity with Figure~\ref{fig:UMAP}  provides evidence in favor of merging them.

\section{Alternative dimensionality-reduction methods}

\begin{figure}
\centering
\includegraphics[width=0.80 \linewidth]{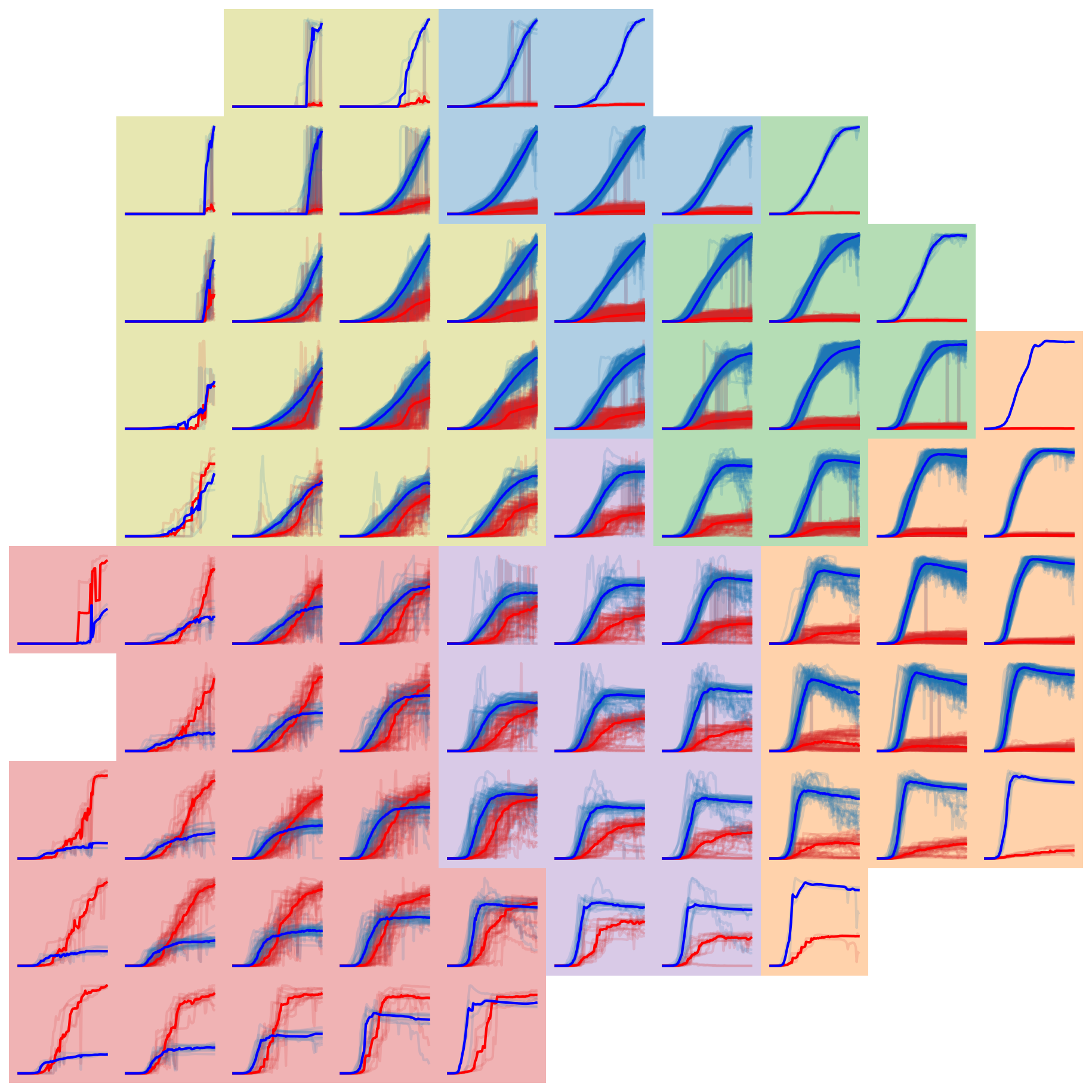}
\includegraphics[width=0.80 \linewidth]{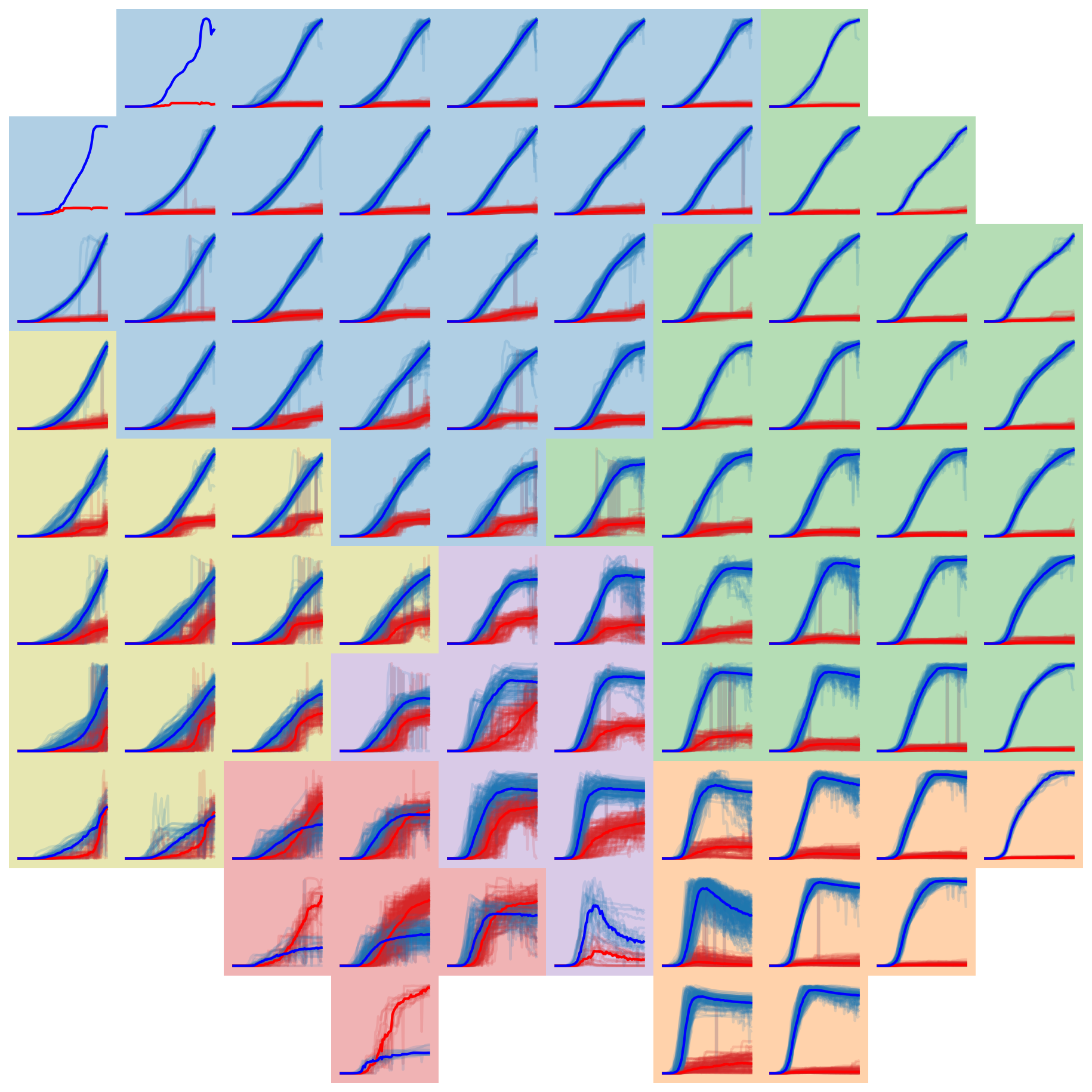}
\caption{\label{fig:2DSFH} Star formation histories in PCA (top) and t-SNE (bottom) representation. Plots are divided using a 10 x 10 grid. Details of each plot are the same as those in Figure~\ref{UMAPSFH}.}
\end{figure}

Figure~\ref{fig:2DSFH} explores alternative dimensionality-reduction methods, namely PCA (top) and t-SNE (bottom), using the same methodology as in the main analysis shown in Figure~\ref{UMAPSFH}. Both methods recover qualitatively similar patterns of star formation and quenching, as well as variations associated with merger activity, across the projected spaces. The cluster structure, indicated by the color coding and obtained using the same $k$-means algorithm in all three representations, is broadly consistent with that found using UMAP, although the locations and relative sizes of the clusters differ slightly between methods. This agreement supports the robustness of our results, suggesting that the main patterns reflect intrinsic properties of the galaxy SFHs rather than being specific to a particular dimensionality-reduction technique. We therefore adopt UMAP as our main method. PCA is a linear technique and may be less effective at capturing the complex, non-linear structure of the SFHs. Although t-SNE can also reveal non-linear structure, it is computationally more expensive in our analysis and does not provide a clear advantage over UMAP for our specific goals.

\section{Clustering alternatives}

\begin{figure*}
\centering
\includegraphics[width=0.40 \linewidth]{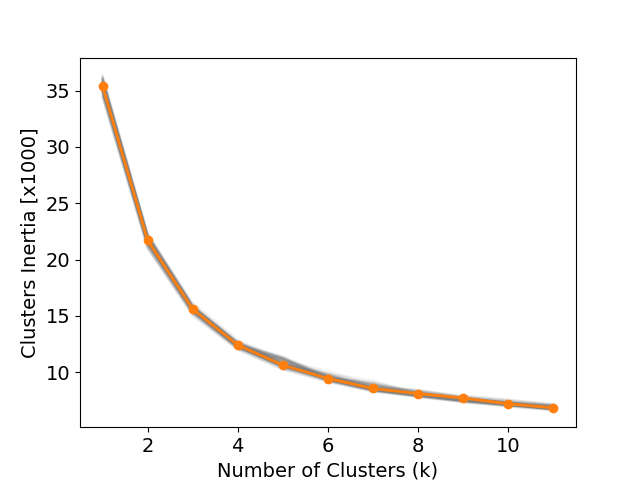}
\includegraphics[width=0.40 \linewidth]{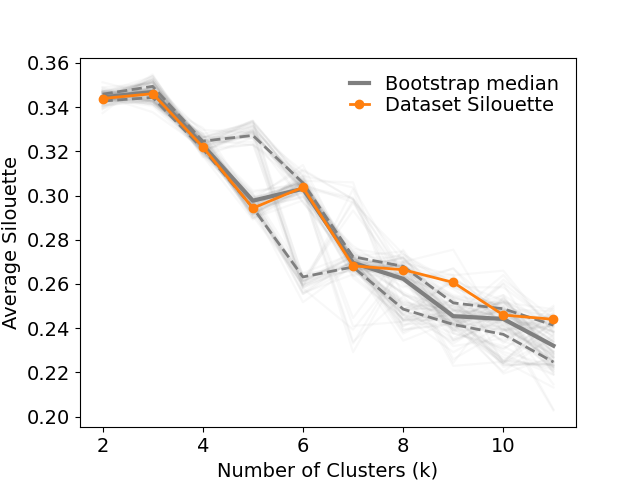}
\caption{\label{fig:ElbowKmeans}Left: Elbow plot for $k$-means clustering. Clusters inertia (within cluster variation) vs the number of clusters. Gray lines correspond to 100 bootstrap resamplings (with replacement) of the whole dataset as stability test. Right: Average silhouette vs number of clusters. Solid gray thick line corresponds to the median and dotted lines correspond to the first and third quartiles.}
\end{figure*}

We present Figure~\ref{fig:ElbowKmeans} with two standard metrics used to determine the number of clusters. On the left side we have the so called elbow plot, where the clusters inertia is displayed versus the number of clusters, indicating that 3 to 6 are reasonable choices (we look for the point where the curve flattens). On the right side we display the mean silhouette coefficient\footnote{For each point $S_i = \frac{b_i-a_i}{\max{(a_i,b_i})}$ where $a_i$ is the mean intra-cluster distance, and $b_i$ is the mean nearest cluster distance.} (that measures how near are observations to their own cluster compared with the distance to the neighbor cluster). The absolute maximum is on 3 (too simple for our analysis, because evolutionary paths would not be detailed enough) but there is also a local maximum at 6 (our choice).

\begin{figure*}
\begin{subfigure}{0.33 \textwidth}
  \centering
  \includegraphics[width=1.00\linewidth]{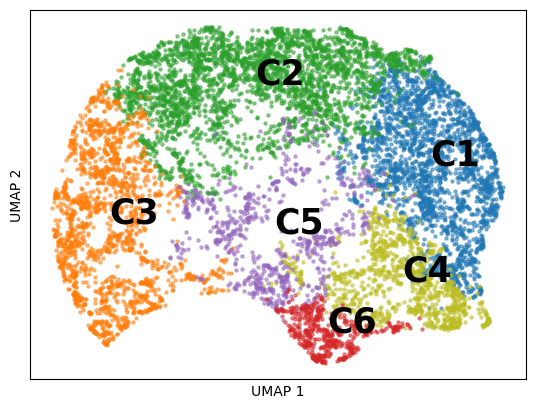}
  \caption{TNG100 and TNG50}
  \label{fig:ClusterWard1}
\end{subfigure}
\begin{subfigure}{0.33 \textwidth}
  \centering
  \includegraphics[width=1.00\linewidth]{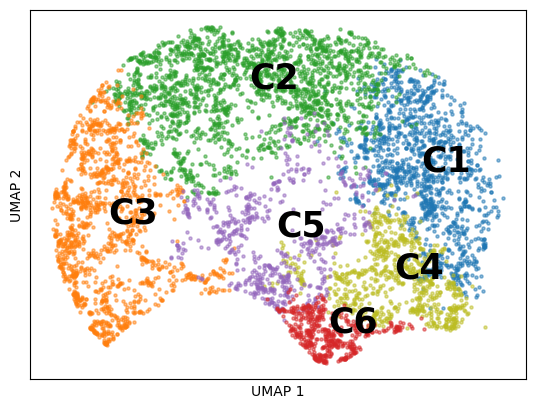}
  \caption{TNG100}
  \label{fig:ClusterWard2}
\end{subfigure}
\begin{subfigure}{0.33 \textwidth}
  \centering
  \includegraphics[width=1.0\linewidth]{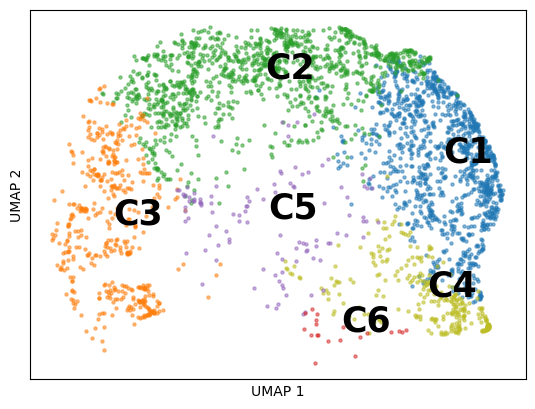}
  \caption{TNG50}
  \label{fig:ClusterWard3}
\end{subfigure}

\caption{Results of agglomerative hierarchical clustering (euclidean metric and Ward linkage) with the number of clusters set to 6. The result is displayed in the UMAP 2D parametric space for the merged data TNG100 $\&$ TNG50 (left), TNG100 (middle) and TNG50 (right) separately.}
\label{fig:ClusterWard}
\end{figure*}

The choice of the clustering method is another important variable, although $k$-means is our main clustering algorithm, other methods have been explored, in particular agglomerative hierarchical clustering is shown in Figure~\ref{fig:ClusterWard} where the euclidean metric and the Ward linkage have been used for better comparability with $k$-means. It uses the same reduction technique (UMAP) and color code that Figure~\ref{fig:ClusterKmeans} but with a different clustering algorithm. The results are virtually the same. Again, regarding our choice, hierarchical clustering is more expensive from a computational point of view and in this case gives slightly more overlapped clusters, so $k$-means is the method used in the main body of our work.

\begin{figure*}
         \centering
         \includegraphics[width=0.33\textwidth]{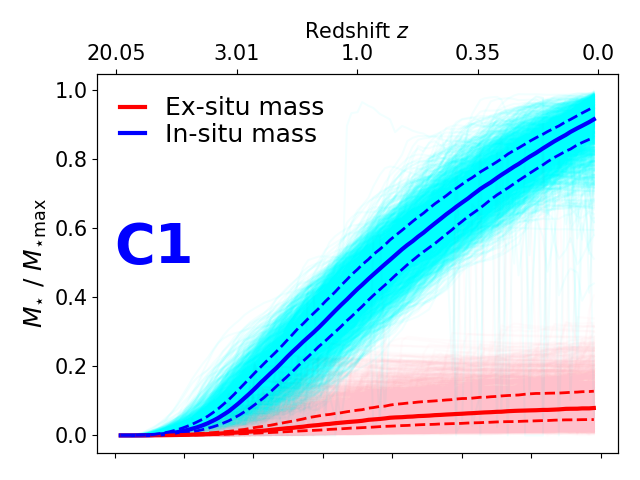}
         \includegraphics[width=0.33\textwidth]{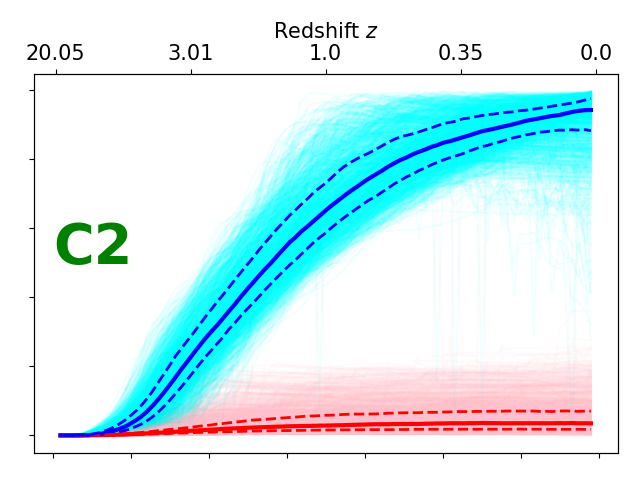}
         \includegraphics[width=0.33\textwidth]{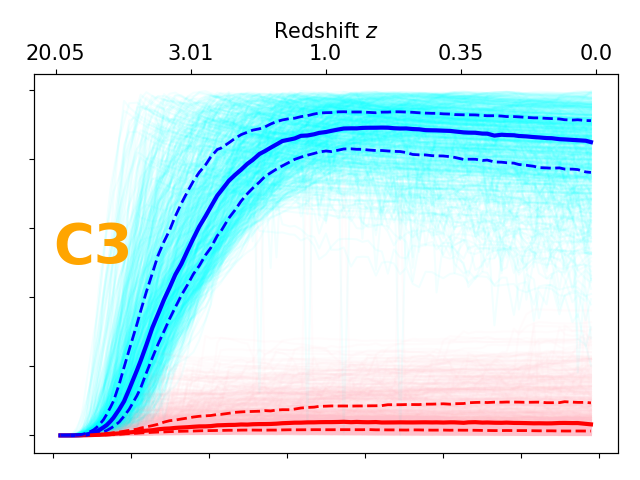}
         \includegraphics[width=0.33\textwidth]{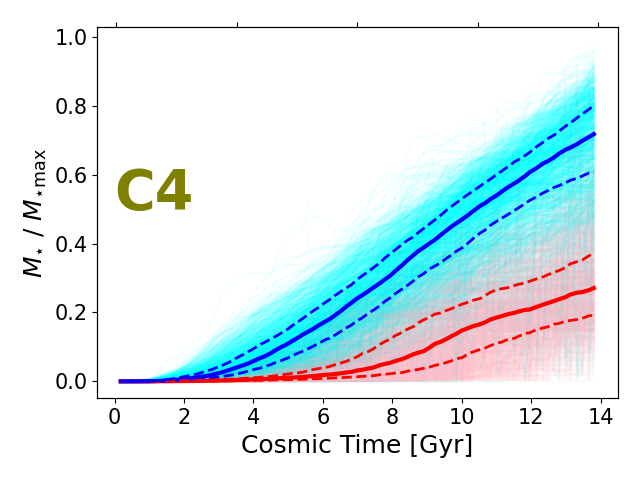}
         \includegraphics[width=0.33\textwidth]{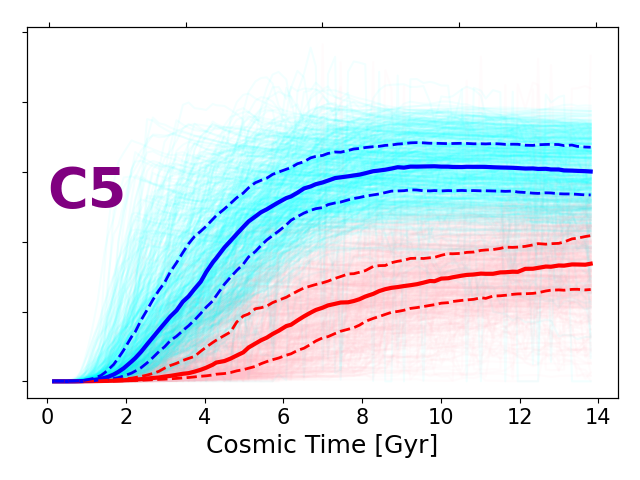}
         \includegraphics[width=0.33\textwidth]{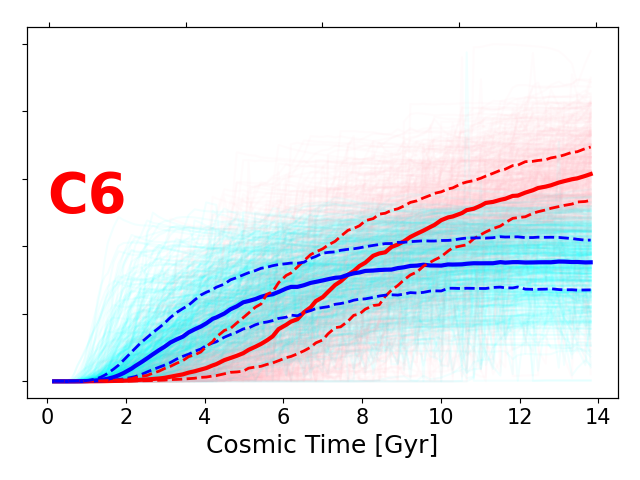}
        \caption{Same as Figure ~\ref{fig:SFHCluster} but using only TNG100 data.}
        \label{fig:SFHCluster100}
\end{figure*}

We also present Figure~\ref{fig:SFHCluster100} where $k$-means clustering is applied only to TNG100 data. The similarity between both Figure~\ref{fig:SFHCluster100} and Figure~\ref{fig:SFHCluster} provides evidence in favor of merging both catalogs.

\end{appendix}

\end{document}